\documentclass{emulateapj}

\usepackage{booktabs}
\usepackage{amssymb}
\usepackage{url}
\usepackage{natbib}
\usepackage[bookmarks]{hyperref}
\usepackage[ps2pdf]{thumbpdf}

\def\sun{\hbox{$_\odot$}}
\newcommand{\HII}{\ion{H}{2}~}

\newcommand{\SiII}{[\ion{Si}{2}]~}
\newcommand{\SI}{[\ion{S}{1}]~}
\newcommand{\SIII}{[\ion{S}{3}]~}
\newcommand{\FeII}{[\ion{Fe}{2}]~}
\newcommand{\NeII}{[\ion{Ne}{2}]~}
\newcommand{\NeIII}{[\ion{Ne}{3}]~}
\newcommand{\PIII}{[\ion{P}{3}]~}

\begin{document}

\title{Infrared spectroscopy of intermediate mass young stellar objects}
\author{Jan Pitann\altaffilmark{1}, Martin Hennemann\altaffilmark{2}, Stephan Birkmann\altaffilmark{3}, Jeroen Bouwman\altaffilmark{1}, Oliver Krause\altaffilmark{1}, and Thomas~Henning\altaffilmark{1}
}
\affil{$^1$Max-Planck-Institut f\"ur Astronomie, K\"onigstuhl 17, D-69117 Heidelberg, Germany, pitann@mpia.de\\
$^2$Laboratoire AIM, CEA/IRFU-CNRS-Universit\'e Paris Diderot, Service d'Astrophysique, CEA Saclay, 91191 Gif-sur-Yvette, France\\
$^3$ESA/ESTEC, Keplerlaan 1, Postbus 299, 2200 AG Noordwijk, Netherlands
}

\shorttitle{IRS spectroscopy of YSO}
\shortauthors{Pitann et al.}
\accepted{by \apj ~2011 August 24}
\submitted{}

\begin{abstract}
In this paper we present \textit{Spitzer} Infrared Spectrograph spectroscopy for 14 intermediate-mass young stellar objects. 
We use \textit{Spitzer} spectroscopy to investigate the physical properties of these sources and their environments. Our sample can be divided into two types of objects: young isolated, embedded objects with spectra that are dominated by ice and silicate absorption bands, and more evolved objects that are dominated by extended emission from polycyclic aromatic hydrocarbons (PAHs) and pure H$_2$ rotational lines. We are able to constrain the illuminating FUV fields by classifying the PAH bands below 9~$\mu$m. For most of the sources we are able to detect several atomic fine structure lines. In particular, the \NeII line appearing in two regions could originate from unresolved photodissociation regions (PDRs) or J-shocks. We relate the identified spectral features to observations obtained from NIR through submillimeter imaging.
The spatial extent of several H$_2$ and PAH bands is matched with morphologies identified in previous \textit{Spitzer}/IRAC observations. This also allows us to distinguish between the different H$_2$ excitation mechanisms. In addition, we calculate the optical extinction from the silicate bands and use this to constrain the spectral energy distribution fit, allowing us to estimate the masses of these YSOs.
\end{abstract}

\section{Introduction} \label{Intro}

Despite significant progress in our understanding of star formation, the earliest phases under which stars form are poorly characterized. This is particularly the case for stars of intermediate to higher mass ($\ge 2\;M_\odot$), for which the evolutionary timescales are much shorter and which are smaller in number, compared to low-mass stars.

Cloud cores that can form intermediate-mass stars have masses of $\sim 10^2-10^3\;M_\odot$ and temperatures of $10-20$~K and appear in most cases as radio-quiet \citep{Arvidsson10}. 
Submillimeter studies of infrared dark clouds (IRDCs) also show high column densities ($10^{22}$~cm$^{-2}$) capable of supporting intermediate to high-mass star-formation \citep{Vasyunina09}.
We have been studying a number of star forming regions, which are in an early evolutionary stage, matching these requirements, based on a large-scale unbiased sample of cold cloud cores or IRDCs (see Section \ref{sec:Tar}).
Within these regions embedded mid-infrared point sources have been identified.
A subsample of these sources are good candidates for intermediate mass young stellar objects (YSO).
To study their physical conditions and chemical characteristics, we conducted infrared spectroscopy on these sources.

One of the first infrared spectroscopic observation of a nearby star-forming region, performed by the \textit{Infrared Space Observatory} (\textit{ISO}), was a study of Orions IRc2 \citep{vanDishoeck98,Wright00}.
Due to the large aperture of \textit{ISO}'s Short Wavelength Spectrograph (SWS), the measured spectrum is a superposition of multiple physical components. The IRc2 spectrum contains hydrogen recombination lines, ionic fine structure lines and broad UV-pumped polycyclic aromatic hydrocarbon (PAH) emission bands originating from a photodissociation region (PDR). Thermally excited, rotational, and rovibrational lines of H$_2$ are also observed and can be used to infer the coupling between the molecular gas and dust grains. The PAHs and the smallest dust grains in this region are photoelectrically heated by a central hot object. These observations show solid state absorption features, such as silicates, water, methane and CO$_2$ ice, originating from a quiescent extended ridge. In addition, shock excited molecular hydrogen was observed for the adjacent Orion peaks I and II \citep{Rosenthal00}. A detailed study of the molecular gas-phase features of CO$_2$, C$_2$H$_2$ and HCN appearing in absorption for IRc2 and to some extent in emission for the Orion peaks can be found in \citet{Boonman03}.

In this work, we use data obtained by the Infrared Spectrograph \citep[IRS;][]{Houck04} on-board the \textit{Spitzer Space Telescope} \citep{Werner04}. We find similar spectral features and use them to estimate the physical properties of the YSOs in the observed star-forming regions. From the 9.7~$\mu$m silicate features the optical extinction can be calculated and used for later spectral energy distribution (SED) fitting (see \citet{Henning10} for a general review on silicates). By classifying PAH bands below 9~$\mu$m it is also possible to constrain some properties of the irradiating FUV field \citep{Peeters02,Tielens08}. Class $\mathcal{A}$ PAH spectra are typical for \HII  and compact \HII  regions, reflection nebula, and YSOs exposed to a similar UV radiation field strength. The typical UV flux intensities are $10^3-10^6\;G_0$.  Referring to \citet{Tielens08}, class $\mathcal{A}$ spectra are excited by sources with $T_\mathrm{eff}>10000$~K. This class of PAH spectra have an asymmetric feature at 6.2~$\mu$m, an emission complex peaking at 7.6~$\mu$m and a third PAH band at 8.6~$\mu$m. These PAH features arise mostly from the C$-$C mode.
Strong UV radiation fields and gas densities of $n \sim 10^3-10^5\;\textrm{cm}^{-3}$ are needed to form photodissociationregions (PDR) associated with molecular gas \citep{Hollenbach97}.
The gas densities and the strength of the UV radiation field can be estimated from the ratios of certain forbidden atomic lines (e.g., \FeII, \SiII) and the excited H$_2$ lines which were created in the deeper layers of the illuminated PDR \citep{Kaufman06}.
It is not only in PDRs that emission lines of ionized atoms can be observed. J-shocks interacting with the envelope of young protostars leave hot dense gas heated up to $10^5$~K.
In the region behind the shock front, the molecules are dissociated and the ionized atoms can be observed by strong infrared emission lines (e.g., \NeII 12.8~$\mu$m, \FeII 26~$\mu$m, \SiII 34.8~$\mu$m, \SI 25~$\mu$m) \citep{Hollenbach89}. The \NeII atomic fine structure line intensity depends on the shock velocity. Therefore strong \NeII lines can be only observed in high velocity J-shocks ($v_s > 60-80$~km$\,$s$^{-1}$). When the gas is cooling in the relaxation region further downstream from the shock front, H$_2$ is formed and can be observed as pure rotational lines. In the non-dissociated C-shocks the temperatures never become high enough to dissociate molecular material. Spectral indications for C-shocks are strong molecular lines of H$_2$, CO, H$_2$O, and OH \citep{Kaufman96}. 
Most spectroscopic studies of massive star-forming regions have used data from \textit{ISO}'s SWS and Long Wavelength Spectrograph (LWS). The IRS on board the \textit{Spitzer Space Telescope} presents a different instrumental approach. It has a lower spectral resolution, but provides a better spatial resolution and an improved sensitivity compared to \textit{ISO}. The long slit spectroscopy in the low resolution mode allows spatially resolved observations of selected lines down to a resolution of 1.8$''$. Nevertheless multiple emission or absorption features from different spatial positions are still covered by the IRS beam. Therefore an interpretation of the spectroscopic results can be given only when supplemented with multi-wavelength imaging observations. In general, these observations have a better spatial resolution than IRS in the same wavelength order and can therefore be used to trace certain spectral features. For example, the bands from the \textit{Spitzer}'s Infrared Array Camera \citep[IRAC;][]{Fazio04} and the Multiband Imaging Photometer \citep[MIPS;][]{Rieke04} can be used to determine the spatial distribution of excited hydrogen, PAHs, and warm dust. 
In particular the 4.5~$\mu$m band is utilized to identify the so-called green and fuzzy features \citep{Chambers09}. These features predominantly originate from the (0-0) $S(9)$ H$_2$ band, mostly attributed to shock excitation \citep[e.g.][]{Buizer10}. Another excitation mechanism for the $S(9)$ line can be fluorescence, depending on physical parameters (e.g. temperature, H$_2$ column densities) and dust properties \citep{Black87}. 
Submillimeter observations (e.g. SCUBA, ATLASGAL) can also be used to trace cold dust. The current observations with Herschel will close the gap between the mid-infrared regime and the ground based submillimeter observations. For the cold ($\sim 20$~K) dust cores found, e.g. in the ISOPHOT 170~$\mu$m serendipity survey \citep[ISOSS; ][]{Bogun96}, 
the Planck spectrum peaks in the Herschel PACS bands.

The next section describes the target selection. In Section \ref{sec:Obs}, we describe the observations and the data reduction processes. Section \ref{sec:res} details the continuum observations and the spectral features. Section \ref{sec:Discuss} discusses the different sources and their morphology within the context of previous observations.

\section{Target selection} \label{sec:Tar}

\citet{Krause04} detail a number of suitable sources detected with ISOSS. Due to its high sensitivity, the survey was able to find sources away from the galactic plane. We selected several regions within this sample based on their distance, luminosity and dust temperature ($12$ K $< T_d \leq 22 $~K). Later \textit{Spitzer} imaging observations identified point sources within some of these regions \citep{Birkmann06,Birkmann07,Hennemann08_20,Hennemann09}.  
We conducted a spectroscopic follow up of these objects. All our targets are resolved bright point sources at MIPS 24~$\mu$m. Based on their mid-infrared colors, these sources were good candidates for intermediate-mass YSOs. The basic parameters of the selected ISOSS targets are listed in Table \ref{tab:tar_selc} (top part). In addition to the coordinates and the distances, the stellar mass estimates and the masses of the associated cold cores are given. In the last column the key references for the particular regions are listed.

The ISOSS J18364$-$0221 East, ISOSS J20153$+$3453, ISOSS J04225$+$5150 and ISOSS J23053$+$5953 regions have extended PAH emission and warm dust components, which can be identified in the \textit{Spitzer}/IRAC and MIPS 24~$\mu$m images, respectively. In the MIPS 24~$\mu$m band ISOSS J18364$-$0221 West appears as an isolated infrared source with weak dust components. 

To investigate additional targets similar to those chosen from the ISOSS sample, we selected 24~$\mu$m sources associated with cold cloud core candidates from \citet{Chambers09} that have IRS spectra available in the Spitzer-SSC archive.
They are associated with IRDCs located in the galactic plane (called ``IRDC sources'' hereafter).
 They are clearly resolved point sources in the MIPS 24~$\mu$m and 70~$\mu$m bands (Table  \ref{tab:tar_selc}, bottom part). Except for G10.70-0.13, 1.2~mm observations show compact cores with  masses ranging from 74 to $301\;M$\sun ~\citep{Rathborne06}. For some of these sources H$_2$O maser emission was observed indicating ongoing star formation processes \citep{Chambers09}. Where available, the SCUBA legacy data \citep{Francesco08} for these regions reveal submillimeter clumps. 
In contrast to the other IRDC sources the G10.70-0.13 sources are not located within the dark filament of the IRDC but show extended PAH emissions such as, for example, the ISOSS~J20153$+$3453 and ISOSS~J04225$+$5150 region.

\section{Observations and data reduction} \label{sec:Obs}

\begin{deluxetable*}{lccclll}[th]
\tablecolumns{7}
\small
\tabletypesize{\footnotesize}
\tablewidth{0pt}
\tablecaption{Sources}
\tablehead{\colhead{Source}	& \colhead{R.A.(J2000)}	   & \colhead{Decl.(J2000)}     & \colhead{Distance$^a$}  & \colhead{${M_\mathrm{core}}^b$ } & \colhead{${M_\mathrm{obj}}^c$ } & \colhead{Key Ref.} \\
								& \colhead{(hh:mm:ss)} & \colhead{(dd:mm:ss)} &	\colhead{(kpc)} & \colhead{$(M\sun)$}	& \colhead{$(M\sun)$}    &  }
\startdata 
	ISOSS 				&				&				&		&			& 			&				\\
	J04225$+$5150 East	& 04:22:32.3	& $+$51:50:30	& 5.5	& -			& 5.4-9.0	&  (BPhD)		\\
	J18364$-$0221 East	& 18:36:36.0	& $-$02:21:59	& 2.2	&  75		& 	-		& (B06), (H09)	\\
	J18364$-$0221 West	& 18:36:29.6	& $-$02:21:59	& 2.2	&  280		& 0.6-2.0 	& (B06)			\\
	J20153$+$3453 East	& 20:15:21.2	& $+$34:53:46	& 2.0	& $87-149$ 	&  	-		& (H08)			\\
	J20153$+$3453 West	& 20:15:20.6	& $+$34:53:53	& 2.0	& $87-149$ 	& 4.0-8.5	&  (H08)		\\
	J23053$+$5953 East	& 23:05:23.6	& $+$59:53:56	& 3.5	& 200		& 4.1-6.6	& (B07)			\\
	\midrule
	G010.70-0.13 I		& 18:09:57.8 	& $-$19:48:17.6	& 3.7	& -			& 5.3-12.0	& (P09) 		\\
	G010.70-0.13 II		& 18:09:55.9 	& $-$19:48:17.6	& 3.7	& -			& 5.6-6.1   & (P09)			\\
	G019.27+0.07		& 18:25:58.6 	& $-$12:03:58.0	& 2.4	& 113		& 2.1-4.9	& (R06), (C09)	\\
	G025.04-0.20		& 18:38:09.6	& $-$07:02:30.0	& 3.4	& 276		& 5.4-9.6	& (R06), (C09) 	\\
	G028.04-0.46		& 18:44:08.2	& $-$04:33:17.5	& 3.2	& 80		& 1.3-7.6	& (R06), (C09)	\\
	G034.43+0.24		& 18:53:20.6	& $+$01:28:24.5	& 3.7	& 301		& 3.6-6.4	& (R05), (R06), (C09) \\
	G053.25+0.04 I		& 19:29:31.9	& $+$18:00:40.0	& 1.9	& -			& 3.6-6.4	& (R06), (C09)	\\
	G053.25+0.04 II		& 19:29:31.7	& $+$18:00:33.2 & 1.9	& -			& 2.3-6.1	& (R06), (C09)	
    \enddata
    \tablecomments{$^\mathrm{a}$ The distances are obtained from the references in the last column. The kinematic distance for G010.70-0.13 is based on CO data from \citet{Dame01}.\\
	$^\mathrm{b}$ The submillimeter core mass given in the key references for this object, if available. The targets in  ISOSSJ20153$+$3453 are separated by just $10\arcsec$, and the submillimeter peak is centered on the western source.\\
	$^\mathrm{c}$ The masses of the point sources were estimated by the SED fitting tool of \citet{Robitaille07} based on the mid-infrared luminosities and the optical extinctions derived from the silicate bands if present. Because of the missing IRAC counterparts, no mass for the central source was modeled for ISOSS~J18364$-$0221 East and ISOSS~J20153$+$3453 East.}
    \tablerefs{ (BPhD) \citet{BirkmannPhD}; (B06) \citet{Birkmann06}; (B07) \citet{Birkmann07}; (C09) \citet{Chambers09}; (H08) \citet{Hennemann08_20}; (H09) \citet{Hennemann09}; (P09) \citet{Peretto09}; (R05) \citet{Rathborne05}; (R06) \citet{Rathborne06} }
    \label{tab:tar_selc}
\end{deluxetable*}

\subsection{Photometry}
We used the IRAC and MIPS observations described in \citet{Birkmann06,Birkmann07} and \citet{Hennemann09,Hennemann08_20} to measure the integrated fluxes of the ISOSS sources. For the IRDC sources we used data from the Galactic Legacy Infrared Mid-Plane Survey Extraordinaire \citep[GLIMPSE;][]{Benjamin03} and the MIPS Inner Galactic Plane
Survey \citep[MIPSGAL;][]{Carey05}.

\begin{figure}[t]
	\centering
	\plotone{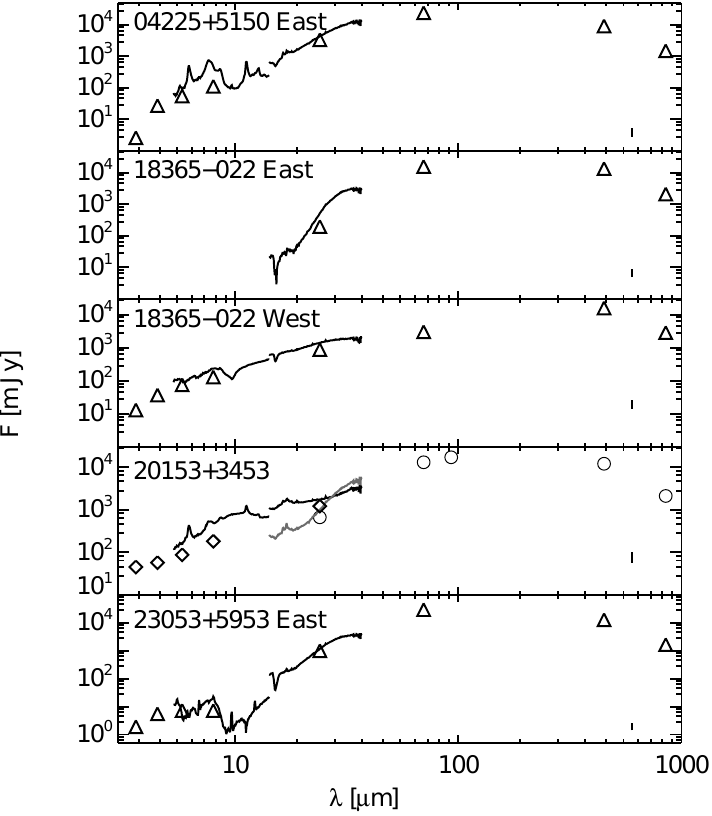}
	\caption{SEDs for the ISOSS targets were compiled using IRAC, MIPS, and SCUBA data. The used apertures for the \textit{Spitzer} photometry are stated in Table \ref{tab:Phot}.
	The bar in the bottom right of each plot indicates an error of 30\% (absolute photometric errors are 7\%, 10\%-20\%, and 20\%-30\%, for IRAC, MIPS, and SCUBA, respectively). The low resolution spectra are overplotted (see also Figures \ref{fig:spec} and \ref{fig:spec_18E_20E}).
	The fourth plot shows the results for ISOSS J20153+3453 East and West.
	Only for the western source are IRAC counterparts detected. The photometric data points (IRAC, MIPS~24~$\mu$m) for the western source are plotted as diamonds, and for the eastern source as circles (MIPS 24~$\mu$m only). Longward of 70~$\mu$m both the sources are not separately resolved. The resulting data points are plotted as circles.}
\label{fig:sed}
\end{figure}

\begin{figure}[ht]
	\centering
	\plotone{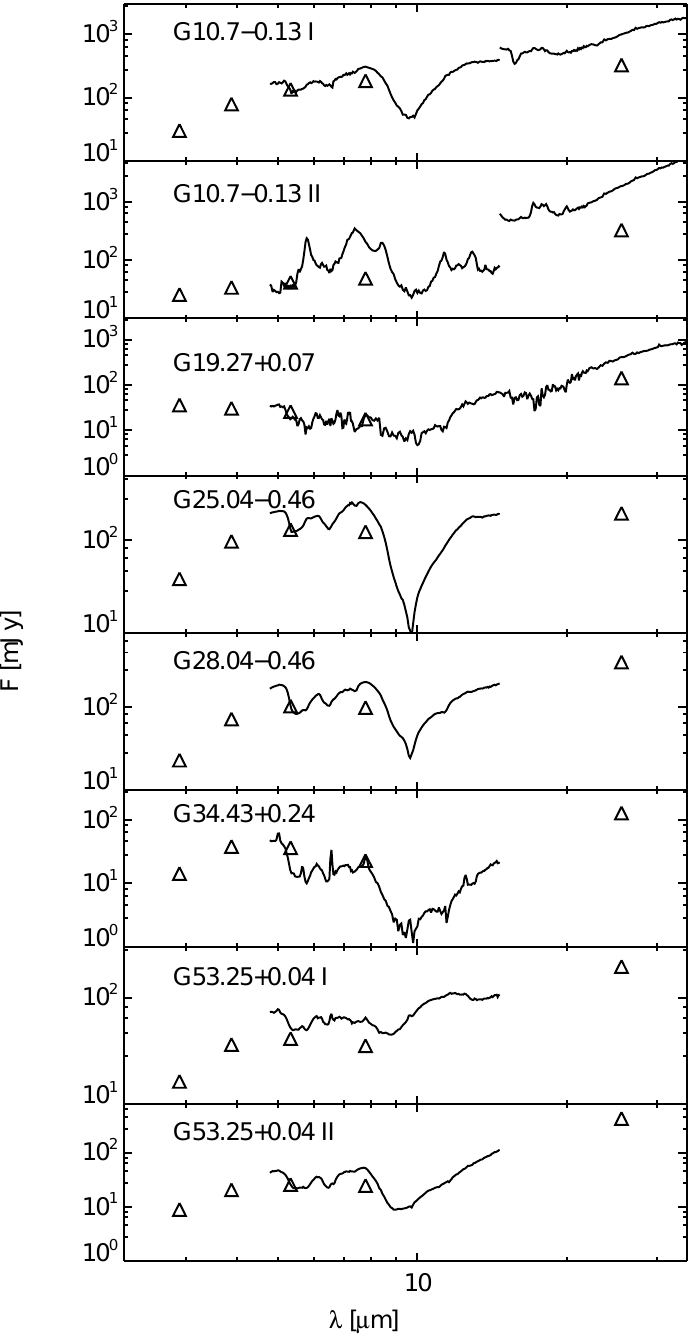}
	\caption{SEDs for the IRDC sources were compiled using IRAC and MIPS data. The bar in the bottom left of each plot indicates the absolute photometric error of $10\%$ for the MIPS 24~$\mu$m band.}
	\label{fig:sed_irdc}
\end{figure}

The \texttt{daophot} package in \texttt{IRAF} was used to perform this aperture photometry. The resulting SEDs are shown in Figure \ref{fig:sed} and \ref{fig:sed_irdc}. The apertures and annuli used for the sky background estimations for the \textit{Spitzer} photometry are given in Table \ref{tab:Phot}. Aperture corrections were not applied to the IRAC fluxes due to the inhomogeneous and asymmetric-extended background emission. 
The initial calibration of the IRAC and MIPS BCD data results in uncertainties of 2\% for IRAC (Reach 2005), 4\% for MIPS 24~$mu$m \citep{Engelbracht07} and 10\% for MIPS 70~$\mu$m \citep{Gordon07}. Cosmetic corrections and astrometric enhancements were performed using the MOPEX software package \citep{Makovoz05}. The final images were combined using the IRAF framework. The resulting accuracy for the aperture photometry is estimated  to be $7\%$ for the IRAC and $10\%$ for the MIPS 24~$\mu$m.
For the MIPS 70~$\mu$m band, we used point PSF photometry to decrease the uncertainties introduced by flux non-linearities \citep{Gordon07}. The resulting uncertainty for MIPS 70~$\mu$m is estimated to be $20\%$.
For the ISOSS J18364$-$0221 East region the MIPS 24~$\mu$m peak was not detected as a point source in the IRAC image and as such photometry was only performed on the MIPS exposures.
IRAC sources were not detected at the position of the eastern submillimeter peak in the ISOSS J20153$+$3453 region. A cluster of IRAC sources were detected at the position of the western 24~$\mu$m source. The photometry was performed on the central object of this cluster; this is also the position on which the IRS slit was centered. The eastern and western targets were resolved at 24~$\mu$m, but could not be detected separately in the MIPS 70~$\mu$m band due to the large beam size. All data points from the eastern and western sources are shown in the same plot (Figure \ref{fig:sed}), and the data points longward of 70~$\mu$m are treated as belonging to both sources. 

450~$\mu$m and 850~$\mu$m observations were obtained with SCUBA \citep{Holland99} at the James Clerk Maxwell telescope (JCMT) for most of the targets by \citet{Birkmann06,Birkmann07} and \citet{Hennemann08_20,Hennemann09}. ISOSS J04225$+$5150 East was not included in these publications, but existing data from 2001 were reduced for the 450~$\mu$m and 850~$\mu$m bands. For ISOSS J04225+5150 we derived submillimeter fluxes of $F_{450}=(9.2\pm 2.8)$~Jy and $F_{850}=(1.5\pm 0.3)$~Jy from the SCUBA bands.
A detailed description of the reduction and analysis process can be found in \citet{Hennemann08_20}. The absolute photometric accuracy is estimated to be 
$30\%$ at 450~$\mu$m and $20\%$ at 850~$\mu$m.

The low resolution spectra are plotted into the SEDs in Figures \ref{fig:sed} and \ref{fig:sed_irdc}.
The differences between spectra and the IRAC/MIPS photometry can be explained as follows: Our sources are compacted  but extended. The available relative spectral response function is only calibrated for point sources (see next the section). 
The spectroscopic observations were conducted with a different aperture compared to the IRAC and MIPS photometries. This results in a significant flux difference. 

\subsection{Infrared spectroscopy}
The spectroscopy was performed with the Infrared Spectrograph (IRS) on board the \textit{Spitzer Space Telescope}. The observations of the targets selected from the ISOSS sample were performed during the campaign ``IRS spectroscopy of extremely young massive proto-stars'' (ID: 30919 + 40569). The observation dates, the program ID, the number of observational cycles and the integration times can be found in Table \ref{tab:obs_log}. 
To cover the maximum possible wavelength interval (5.2~$\mu$m $-$ 38.0~$\mu$m) all four available low resolution channels (SL2, $5.2-7.7$~$\mu$m; SL1, $7.4-14.5$~$\mu$m; LL2, $14.0-21.3$~$\mu$m and LL1, $19.5-38.0$~$\mu$m) and both high resolution channels (SH, $9.9-19.6$~$\mu$m; LH, $18.7-37.2$~$\mu$m) were used. ISOSS J18364$-$0221 East was the only exception; here the SL2 channel was skipped due to the non-detection of a point source in the IRAC bands. The integration times for the low resolution modes were chosen to achieve a signal-to-noise ratio (S/N) of ~100, sufficient to detect faint absorption features. In order to detect faint lines above the relatively bright mid-infrared continuum, we also aimed for a signal to noise ratio of at least 100 in the LH module at 24~$\mu$m. The integration times for the SH module were optimized in order to reach a spectral line sensitivity for H$_2$ $S(1)$ at $\lambda = 17.03$~$\mu$m, which is two times better than that of the $S(0)$ line. Overall we aimed for a line sensitivity better than $3\times 10^{-18}$~Wm$^{-2}$ and $5\times 10^{-18}$~Wm$^{-2}$ for the short and long wavelength orders of the high resolution spectra. All Astronomical Observation Templates (AOTs) were performed in the standard staring mode. The science targets were placed at 1/3 and 2/3 of the slit length to obtain spectra at two nodded positions. The slit overlays are shown for one nod position in Figures \ref{fig:slit_20EW}, \ref{fig:slit_04E}, \ref{fig:slit_G10},  \ref{fig:slit_18E}, \ref{fig:slit_23E}, \ref{fig:slit_18W}, and \ref{fig:slit_IRDC}. The slit length covers several ten to hundred thousand AU, as indicated in those plots. 

The IRDC sources G025.04-0.20, G028.04-0.46, G034.43+0.24 and G053.25+0.04~I+II were taken in the short wavelength orders only. The observations were carried out with a minimum of 10 cycles. The good sampling results in a high S/N.

\subsubsection{Low resolution data extraction}

\begin{deluxetable*}{lccrrrrl}[th]
\tabletypesize{\footnotesize}
\tablecolumns{7}
\small
\tablewidth{0pt}
\tablecaption{Observation Log}
\tablehead{\colhead{Source} 	& \colhead{AOR-ID} & \colhead{Obs. Date} & \multicolumn{4}{c}{t$_\mathtt{ramp}$ (s), (cycles)} & Comments on \\
\cmidrule(lr){4-7}
							& 				   &	 (yyyy-mm-dd)	     & \colhead{SL}	& \colhead{LL}	& \colhead{SH}	& \colhead{LH} 	& Data Quality}
\startdata 
	ISOSS				&			&					&			&			&			&			&    \\
	J04225$+$5150 East	& 19235328	& 2007 Mar 15  		&  6,(2) 	&  6,(2)	& 30,(2) 	& 14,(2)	&    \\
	J18364$-$0221 East	& 23071744	& 2008 Apr 26  		& 60,(3) 	& 30,(2) 	& 120,(6) 	& 60,(8) 	& SL2 not observed, bright background emission, \\
						& 			& 	  				&  	 	&  	 	& 	 	& 	 					& SL orders not extracted  \\
	J18364$-$0221 West	& 23072000	& 2008 Apr 26  		&  6,(2) 	&  6,(2) 	& 120,(2) 	& 60,(2) 	&    \\
	J20153$+$3453 East	& 19236864	& 2006 Jul 04  		& 14,(4) 	& 14,(2) 	& 120,(3) 	& 60,(4) 	& The western source dominates SL orders \\
	J20153$+$3453 West  & 19237376	& 2006 Jul 4  		&  6,(2) 	&  6,(2) 	&  30,(5) 	& 14,(5) 	&    \\
	J23053$+$5953 East	& 19237888	& 2006 Jul 31  		& 60,(3) 	&  6,(2) 	& 120,(2) 	& 60,(2) 	& Flux leakage in SL orders \\
						& 			& 	  				&  	 	&  	 	& 	 	& 	 					& Extended PAH band$^\mathrm{a}$\\
	\midrule
	G010.70-0.13 I		& 16943872	& 2006 May 1		&  6,(1) 	&  6,(1) 	& 6,(1)   	& 14,(1) 	&    \\
	G010.70-0.13 II		& 16943872	& 2006 May 01		&  6,(1) 	&  6,(1) 	& 6,(1)   	& 14,(1) 	&    \\
	G019.27+0.07		& 12091648	& 2005 Apr 18		& 60,(2) 	& 14,(2) 	& -	    	& -			& very non-uniform background$^\mathrm{b}$  \\
	G025.04-0.20		& 17823744	& 2007 May 08		& 60,(10) 	& -	 	& -	    	& -		&    \\
	G028.04-0.46		& 17823488	& 2007 May 02		& 60,(10) 	& -	 	& -	    	& -		&   extended PAH bands$^\mathrm{c}$ \\
	G034.43+0.24		& 17824512	& 2007 May 02 		& 14,(34) 	& -	 	& -	    	& -		&    extended PAH band$^\mathrm{a}$ \\
	G053.25+0.04 I		& 17825280	& 2007 Jun 14		& 60,(10) 	& -	 	& -	    	& -		&    extended PAH band$^\mathrm{a}$\\
	G053.25+0.04 II		& 17825536	& 2007 Jun 14		& 60,(10) 	& - 	 	& -	    	& -		& extended PAH bands$^\mathrm{c}$  
    \enddata
    \tablecomments{ $^\mathrm{a}$ The 11.2~$\mu$m PAH band is spatially extended over the slit. A polynomial fit is used to estimate the background contribution in this wavelength range.\\
	$^\mathrm{b}$  Because of the very non-uniform background the whole spectrum was extracted with a polynomial background fit. \\
    $^\mathrm{c}$ The 11.2~$\mu$m and 12.3~$\mu$m PAH bands are spatially extended over the slit. A polynomial fit is used to estimate the background. A polynomial fit is used to estimate the background contribution in this wavelength range. }
    \label{tab:obs_log}
\end{deluxetable*}

\begin{deluxetable}{ccccc}[hb]
\tabletypesize{\footnotesize}
\tablecolumns{5} 
\tablewidth{0pc} 
\small
\tablecaption{Photometry Apertures for IRAC and MIPS}
\tablehead{\colhead{Source} 	& \multicolumn{2}{c}{IRAC} 					& \multicolumn{2}{c}{MIPS} \\
\cmidrule(lr){2-3}\cmidrule(lr){4-5}
				 				& \colhead{$R_\mathrm{Ap}$} &	\colhead{$R_\mathrm{Sky}$}	& \colhead{$R_\mathrm{Ap}$}	& \colhead{$R_\mathrm{Sky}$}}
	\startdata
	ISOSS			   &	 &	   &	 &		\\
	J04225$+$5150 East & 2$\arcsec$ & 6$\arcsec$ & 5$\arcsec$ & 13$\arcsec$ \\ 
	J18365$-$022 East  & - & - & 6$\arcsec$ & 13$\arcsec$ \\ 
	J18365$-$022 West  & 3.2$\arcsec$ & 5.2$\arcsec$ & 13.2$\arcsec$ & 30$\arcsec$ \\ 
	J20153$+$3435$^{\;\mathrm{a}}$ & 2$\arcsec$ & 6$\arcsec$ & 6$\arcsec$ & 13$\arcsec$ \\ 
	J23053$+$5953      & 2$\arcsec$ & 6$\arcsec$ & 6$\arcsec$ & 13$\arcsec$ \\ 
	\hline 	
	G10.7$-$0.13~I    & 4\farcs8	& 10\farcs8 & 10\farcs4 	& 31\farcs2 \\ 
	G10.7$-$0.13~II   & 4\farcs8	& 10\farcs8 & 10\farcs4 	& 31\farcs2 \\ 
	G19.27$+$0.07     & 4\farcs8	& 12\farcs0 & 10\farcs4 	& 31\farcs2 \\ 
	G25.04$-$0.20     & 7\farcs2	& 14\farcs4 & 10\farcs4 	& 31\farcs2 \\ 
	G28.04$-$0.46     & 7\farcs2	& 14\farcs4	& 10\farcs4 	& 31\farcs2 \\ 
	G34.43$+$0.24     & 7\farcs2	& 14\farcs4	& 10\farcs4 	& 31\farcs2 \\ 
	G53.25$+$0.04~I   & 3\farcs6	& 19\farcs6	& 10\farcs4 	& 31\farcs2 \\ 
	G53.25$+$0.04~II  & 3\farcs6 	& 19\farcs6	& 10\farcs4 	& 31\farcs2
	\enddata
	\tablecomments{$^\mathrm{a}$ Only the western source is detected within the IRAC bands. For MIPS at 24~$\mu$m the same aperture size was used for the eastern and western sources.}
    \label{tab:Phot}
\end{deluxetable}

The low resolution data products were obtained in the \texttt{droopres} state processed by the \textit{Spitzer} pipeline. 
For further data reduction a pipeline based partly on the SMART-package together with the spectral extraction tools from the FEPS \textit{Spitzer} science legacy team were used. A detailed description can be found in \citet{Bouwman06} and \citet{Swain08}. During the reduction process the data products were flatfield and straylight corrected. The correction of rogue and bad pixels was done with a median filter (based on the \texttt{irsclean} routine) and visual inspection.
The spectral extraction was done with a fixed 6 pixel wide aperture for the short-low (SL) orders and 5 pixel  wide aperture for the long-low (LL) orders, referring to spatial widths of $10.8''$ and $25.5''$ in the slit respectively.
The local background was subtracted using the two nod positions along the slit from a median combined series of exposures. After the extraction the spectrum was convolved with the relative spectral response function (RSRF). The RSRF for the IRS instrument is only defined for point sources. Due to the compact, but extended nature of the sources a proper absolute flux calibration cannot be achieved for the one-dimensional spectra. Since there is a flux offset between the SL and LL orders and the corresponding slits have different orientations, the one-dimensional spectra are shown in two separate plots in Figure \ref{fig:spec}. Furthermore, the fluxes are given in units of Jy with respect to the used aperture sizes.\\

\subsubsection{High resolution data extraction}
The high resolution data (SH: short-high orders; LH: long-high orders) were reduced with the c2d pipeline \citep{Lahuis06,Lahuis10} starting with the \texttt{rsc} data products. Corrections for bad and rogue pixels were performed, as was defringing for all spectral orders. The optimal spectral extractions were obtained using analytically derived cross-dispersion profiles that fit the source profile and extended emission components. The background was determined from additional off-source AOTs for every source. We used the high-resolution data to obtain line fluxes and flux ratios.\\

\subsubsection{Limitations on the extraction process}

\begin{figure*}[ht]
	\centering
	\includegraphics[width=0.55\textwidth]{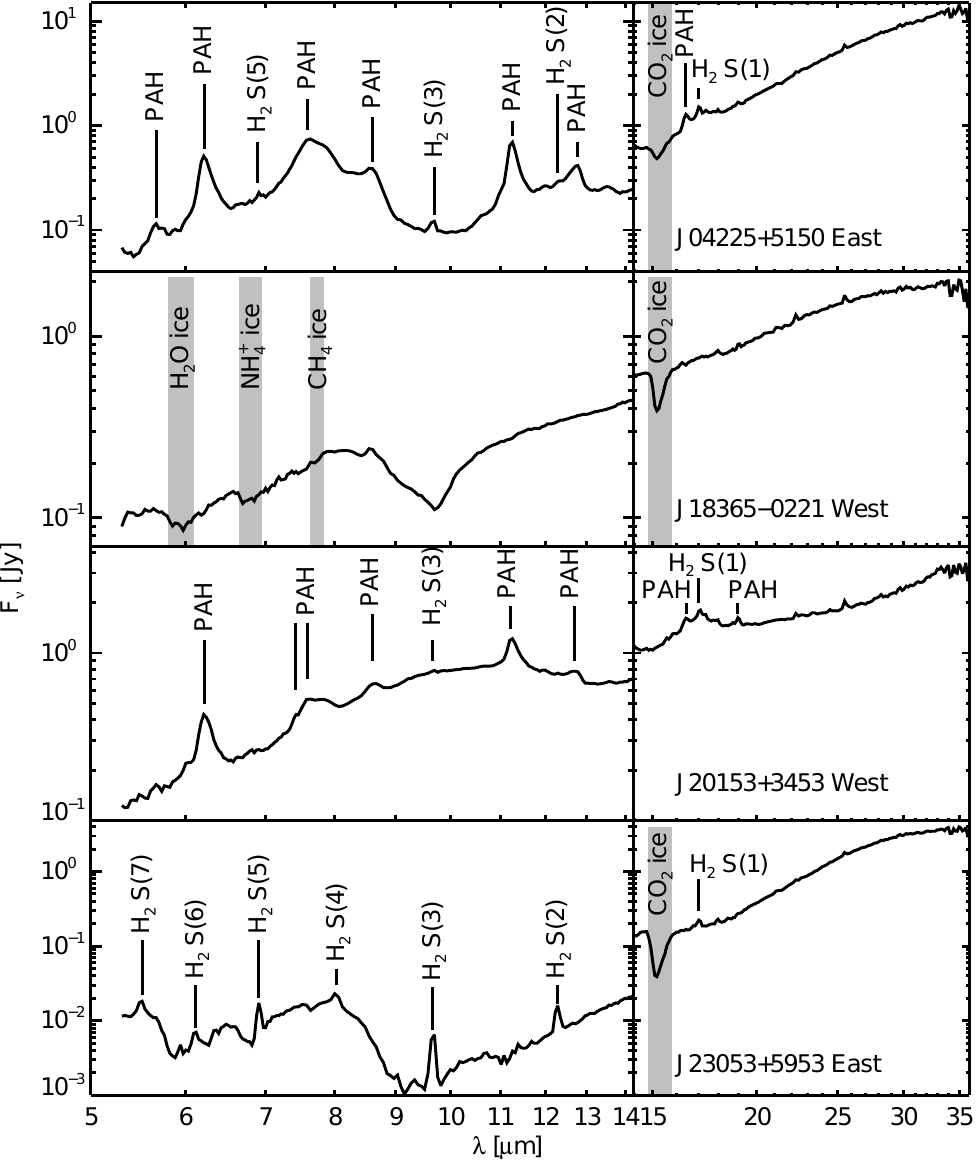}
	\caption{Low-resolution spectra for the ISOSS sources. 
The SL and LL orders are extracted with different aperture size. Therefore the long wavelength orders are more affected by contributions of extended emissions and show a clear offset compared to the short wavelengths.
The low resolution, long wavelength orders for ISOSS~J18365-0221~East and ISOSS~J20153+3453~East are shown in Figure \ref{fig:spec_18E_20E}. Emission lines are indicated by lines, while absorption bands are underlined in light gray.}
	\label{fig:spec}
\end{figure*}

To analyze the effect of extended emission lines and an inhomogeneous background on the low-resolution spectra, we investigated the dependence of the sources' cross-dispersion profiles as a function of wavelength.
Several spectra showed extended spatial profiles, in particular for the 11.3~$\mu$m and 12.8~$\mu$m PAH bands. These extended PAH background contributions could lead to a background over-subtraction in the resulting low resolution spectra. We compared different extraction methods for the low resolution spectra. We used a polynomial background fit for the FEPS extraction in the affected wavelength regimes. The second method was the PSF-extraction from the c2d pipeline, which is slightly undersampled for the low resolution case. Both extraction methods do not significantly differ in the resulting overall continuum, correcting for extended background line emissions. For some sources the c2d extraction returns a slightly better S/N for the continuum. In some cases the c2d pipeline did not detect some of the faint H$_2$ rotational lines. In general the performance of fitting the SL order is better with the FEPS extraction. Based on these facts we chose to process sources which were affected by extended PAH emissions with the nominal FEPS extraction and substitute the affected wavelength intervals with the polynomial background fit (see Table \ref{tab:obs_log}).

ISOSS J18364$-$0221 East shows bright and non-uniform background emissions at short wavelengths ($5.2-14.5''$). No source profile could be fit for these wavelength orders and as such no SL spectrum is presented in Figure~\ref{fig:spec}.
For the ISOSS J20153$+$3453 region two targets were observed centered on the two MIPS peaks at 24~$\mu$m. The separation of both MIPS peaks at 24~$\mu$m is $10''$. The slit positions of the SL orders overlap for the AOTs taken on both these targets. For the eastern target in this region we did not detect a point source in the IRAC bands, but instead detect a diffuse bulge of emission. 
The independent source profile could not be obtained for the SL orders of the eastern target since the slit is dominated by the western source. The short orders were therefore not extracted for the one-dimensional spectrum. The long wavelength orders (LL) are not affected as these slit positions do not overlap.

The high-resolution spectrum could also not be extracted for this target in the SH orders as the cross dispersion profile could not be fitted with the absence of a point source in SH orders.
For J23053$+$5953 East the slit was centered on the peak of the submillimeter clump. The IRAC point source at this position was only partly covered by the slit, which results in a significant flux loss. We can therefore only qualitatively discuss the spectrum below 14.5~$\mu$m.
G019.27+0.07 is affected by a very non-uniform background with bright background emissions. The whole spectrum was extracted with third- and fourth-order polynomial background extractions for the SL and LL orders, respectively. It is shown in Figure \ref{fig:spec_G19} together with the averaged background spectrum at the nominal nod position. 

\subsubsection{Spatial line analysis}
We also investigated the spatial extent of single lines over the low resolution slits (SL,LL). 
A one-dimensional spectrum was extracted for every spatial pixel position over the slit width. This extraction was performed without a background subtraction. The spectra for the two nod positions were extracted separately and were not median combined. The line flux was calculated from a fitted Gaussian profile after the underlying continuum contribution had been removed using a second-order polynomial. These results are presented in Section \ref{sec:res_spatial_extend}.

\begin{deluxetable*}{lccclcll}[hb]
\tabletypesize{\footnotesize}
\tablecolumns{8} 
\tablewidth{0pc} 
\small

\tablecaption{Physical parameters and Spectral features}
\tablehead{\colhead{Source} & \colhead{Distance}  & \colhead{L$_\mathrm{bol}$ $^a$} & \colhead{PAH} &	\colhead{Ices \& Molecular absorptions}	& \colhead{Silicate}	& \colhead{H$_2^{\;\,b}$}	& \colhead{Atomic lines$^c$} \\
							& \colhead{$[$kpc$]$} & \colhead{$[$L\sun$]$}			&				&											&						&					& }
\startdata
	IRc2$^\mathrm{(A)}$			& 0.45				&									& $\checkmark$  &	H$_2$O, CO$_2$, C$_2$H$_2$				& $\checkmark$			&  $\checkmark$		& \NeII, \NeIII, \PIII,  \\
				& &			&       &  &		&		& \FeII, \SI, \SIII, \SiII	\\
\midrule
	ISOSS 		& &			&       &  & 	&		&  \\
	J04225$+$5150 East	& 5.5$^\mathrm{(B)}$	& 5000		& $\checkmark$ &	CO$_2$		& $\checkmark$	& S(0)-S(4), S(5)	& \NeII, \FeII, \SiII \\
	J18364$-$0221 East	& 2.2	&	$70^\mathrm{d}$	        & $\checkmark$ &	-				& -				& S(0)	& \SI, \SiII, \FeII \\
	J18364$-$0221 West	& 2.2	&	550	        & -	   & CO$_2$, H$_2$O, CH$_4$& $\checkmark$		&  -		 & -		\\
	J20153$+$3453 East	& 2.0	&	920	        & $\checkmark$ &	CO$_2$		& $\checkmark$		& S(1)		 & \FeII, \SiII \\
	J20153$+$3453 West	& 2.0	&	920	        & $\checkmark$ &	-		& emiss.				& S(1), S(2) & \FeII, \SiII \\
	J23053$+$5953 East	& 3.5	&	2300        & $\checkmark$ &	CO$_2$		& $\checkmark$		& S(0)-S(7)	 & \FeII, \SiII, \SI \\
	\midrule
	G010.70-0.13 I		& 3.7	&	$39^\mathrm{e}$		& $\checkmark$ & H$_2$O, CO$_2$		& $\checkmark$		& S(1)		 & \NeII, \SiII	\\
	G010.70-0.13 II		& 3.7	&	$18^\mathrm{e}$		& $\checkmark$ &	-		& $\checkmark$				& S(1)		 & \NeII, \SiII	\\
	G019.27+0.07		& 2.4$^\mathrm{(C)}$ &	$9^\mathrm{e}$		& $\checkmark^\mathrm{f}$ &	CO$_2^{\,\;\mathrm{f}}$		& $\checkmark^\mathrm{f}$		& S(0)$^\mathrm{f}$, S(1)$^\mathrm{f}$	 & no hi-res	\\
	G025.04-0.20		& 3.4$^\mathrm{(C)}$ &	$41^\mathrm{e}$		& -	       & H$_2$O, NH$_4^+$, CH$_4$ & $\checkmark$	& -	& 	no hi-res	\\
	G028.04-0.46	& 3.2$^\mathrm{(C)}$ &	$39^\mathrm{e}$		& $\checkmark$ & H$_2$O, NH$_4^+$, CH$_4$ &	$\checkmark$ & S(1)	 & no hi-res	\\ 
	G034.43+0.24	& 3.7$^\mathrm{(D)}$ &	$25^\mathrm{e}$ 	& $\checkmark$ & H$_2$O, NH$_4^+$ 	& $\checkmark$	& S(2), S(4)-S(7) & no hi-res		\\
	G053.25+0.04 I		& 1.9$^\mathrm{(C)}$ &	$5^\mathrm{e}$		& $\checkmark$ & H$_2$O, NH$_4^+$ 	& $\checkmark$	& S(2)-S(7)	& no hi-res	\\
	G053.25+0.04 II		& 1.9$^\mathrm{(C)}$ &	$4^\mathrm{e}$		& $\checkmark$ & H$_2$O, NH$_4^+$ 	& $\checkmark$	& S(4), S(5)	& no hi-res
	\enddata
	\tablecomments{ \\
	$^\mathrm{a}$ The bolometric luminosities were estimated using the SEDs shown in Figure \ref{fig:sed}. \\
	$^\mathrm{b}$ The presence of pure rotational lines in the high or low resolution spectra are listed in this column.\\
	$^\mathrm{c}$ Atomic lines which could not be observed due to missing high resolution spectra, were labeled as not available (no hi-res).\\
	$^\mathrm{d}$ Due to the missing IRAC counterparts the calculated luminosity for ISOSS~J18364$-$0221 East represents just an lower limit. \\
	$^\mathrm{e}$ The luminosities for the IRDC sources are calculated with the IRAC and MIPS bands only.  \\
	$^\mathrm{f}$ Only observed in the background spectrum.}
	\tablerefs{(A) \citet{vanDishoeck98}; (B) \citet{BirkmannPhD}; (C) \citet{Rathborne06}} 
    \label{tab:Lbol}
\end{deluxetable*}

\section{Results} \label{sec:res}

\subsection{Continuum observations and spectral energy distribution}
The SEDs were compiled from IRAC, MIPS and SCUBA bands and are presented in Figure \ref{fig:sed}. 

Using these data we calculated the bolometric luminosities using:
$$L=4\pi d^2 \int d\nu F_\nu$$
This assumes a spherically symmetric distribution
of the luminosity. Of course, other non-spherical configurations are
possible; for example, the majority of the observed luminosity escapes
via outflow cavities. For this reason, the values presented in Table
\ref{tab:Lbol} are affected by large uncertainties.

\begin{figure*}[htp]
	\centering
	\includegraphics[width=0.6\textwidth]{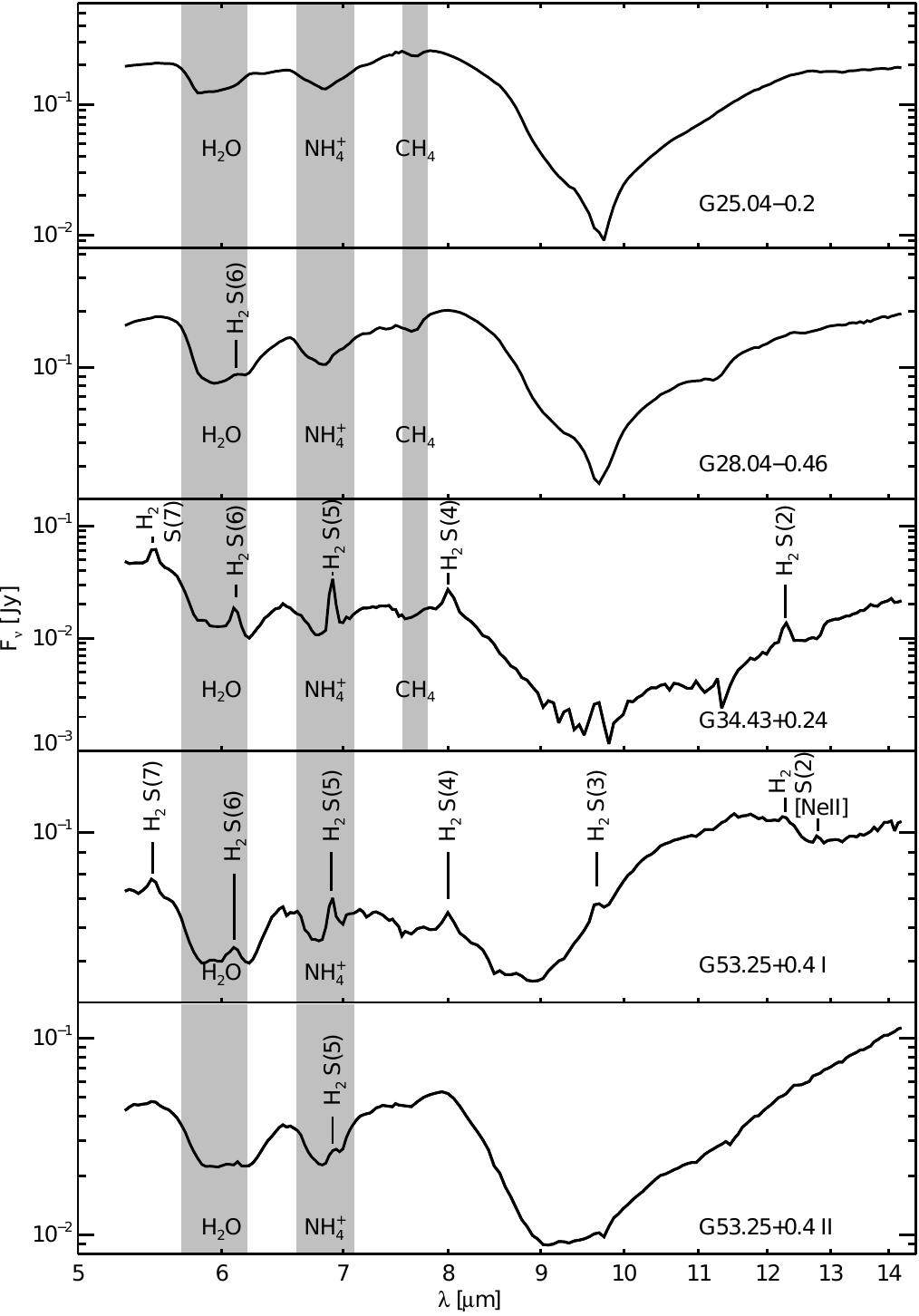}
	\caption{Low-resolution spectra for the IRDC sources. The spectra for G10.7$-$0.13 and G19.27$+$0.07 are shown in Figures \ref{fig:spec_G10} and \ref{fig:spec_G19}.}
	\label{fig:compspec}
\end{figure*}

The low resolution spectra extracted from the ISOSSsources are presented in Figure \ref{fig:spec}. The spectra from the IRDC sources and the long wavelength orders from ISOSS J23053+5953 East and ISOSS J18364$-$0221 East are shown in Figure \ref{fig:compspec}.
Most spectra in our sample are dominated by broad emission bands at shorter wavelengths. Beyond 18~$\mu$m the continuum is well defined and shows a steeper slope  compared to the shorter wavelengths. 
For ISOSS J23053+5953 East the IRAC observations reveal a cluster of objects that are not resolved at the longer wavelengths. The contribution from the clustered sources can be seen in the higher fluxes below 9$\mu$m. Clustering can also be seen in ISOSS J04225+5150 and ISOSS J20153+3453 West.

\subsection{Hydrogen emission} 
\begin{figure}[ht]
	\centering
	\plotone{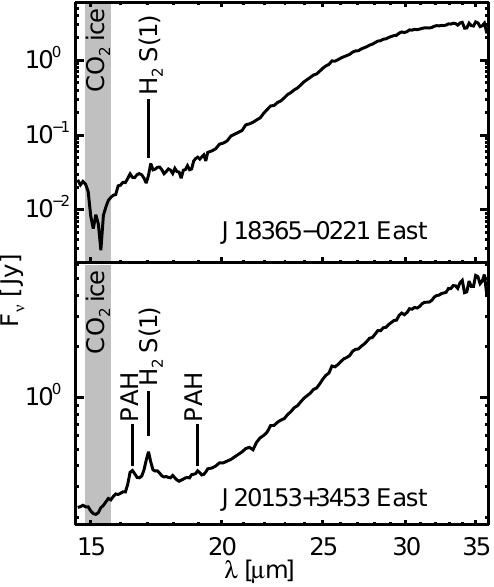}
	\caption{Low-resolution spectra for ISOSS~J18365 East and J20153$+$3453 East are shown. Only the long wavelength orders (LL) were extracted for these sources and plotted.}
	\label{fig:spec_18E_20E}
\end{figure}

The pure rotational lines $(v'-v'')=(0-0)$ detected in our sample range from $S(7)$ to $S(0)$, with corresponding excitation energies of $E_{S_0}/k=510$~K to $E_{S_7}/k=7197$~K \citep[e.g.,][]{Rosenthal00}. 

For 12 out of 14 sources we observed rotational H$_2$ lines. Besides thermal excitation, other mechanisms such as shock excitation or PDR contributions at the atomic-to-molecular interface must be considered (see Section \ref{sec:Discuss}).
Except for three IRDC sources, the low $J<2$ lines with low excitation energies $\leq 1015$~K which can trace warm gas were detected. Only for one source were all the (0-0) transitions present, from $S(0)$ to $S(7)$.
Due to the low S/N ratio in the low resolution spectra the $S(0)$ line can only be identified in the high resolution spectra. The observed rotational lines are listed in Table \ref{tab:Lbol}.
The high resolution spectra reveal the presence of the $S(2)$ (1-1) line for all ISOSS sources apart from the ISOSS J20153+3453 targets.  

For all targets with detected $S(2)$, $S(1)$ and $S(0)$ lines, the line ratios $S(2)/S(1)$, $S(2)/S(0)$ and $S(1)/S(0)$ are below 1.

\begin{figure*}[htp]
	\centering
	\includegraphics[width=0.6\textwidth]{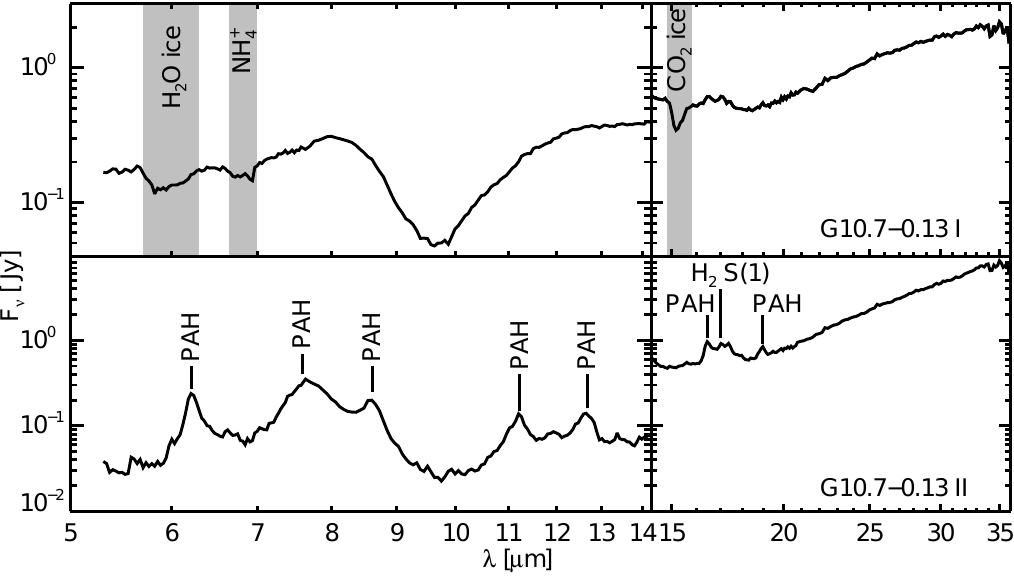}
	\caption{Low-resolution spectra for G10.7$-$0.13~ I and II are shown. }
	\label{fig:spec_G10}
\end{figure*}

\subsection{PAH features}

\begin{figure*}[t]
 \begin{minipage}[b]{.3\linewidth}
    \centering
    \includegraphics[width=1\textwidth]{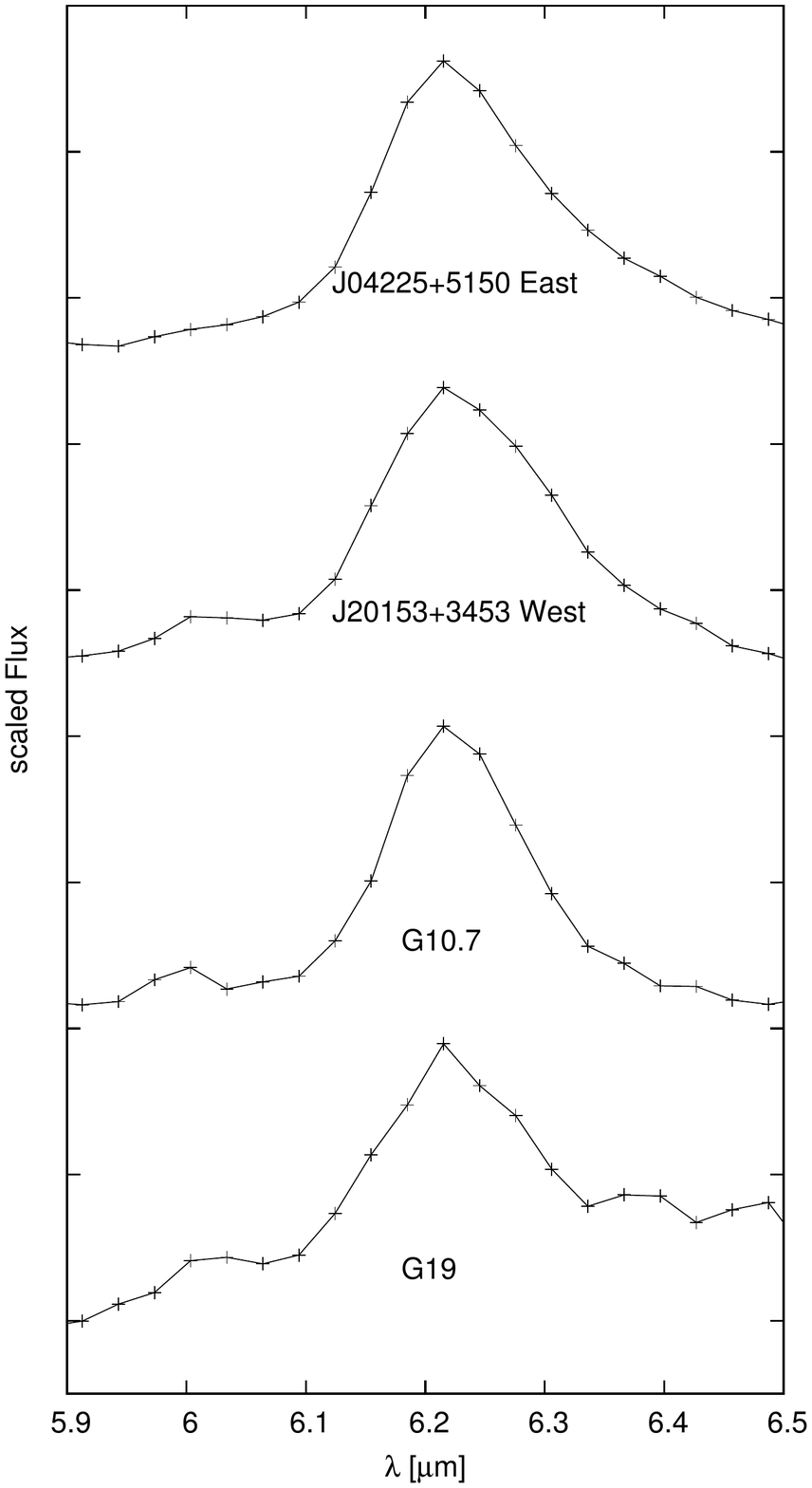}
 \end{minipage}
 \begin{minipage}[b]{.3\linewidth}
    \includegraphics[width=1\textwidth]{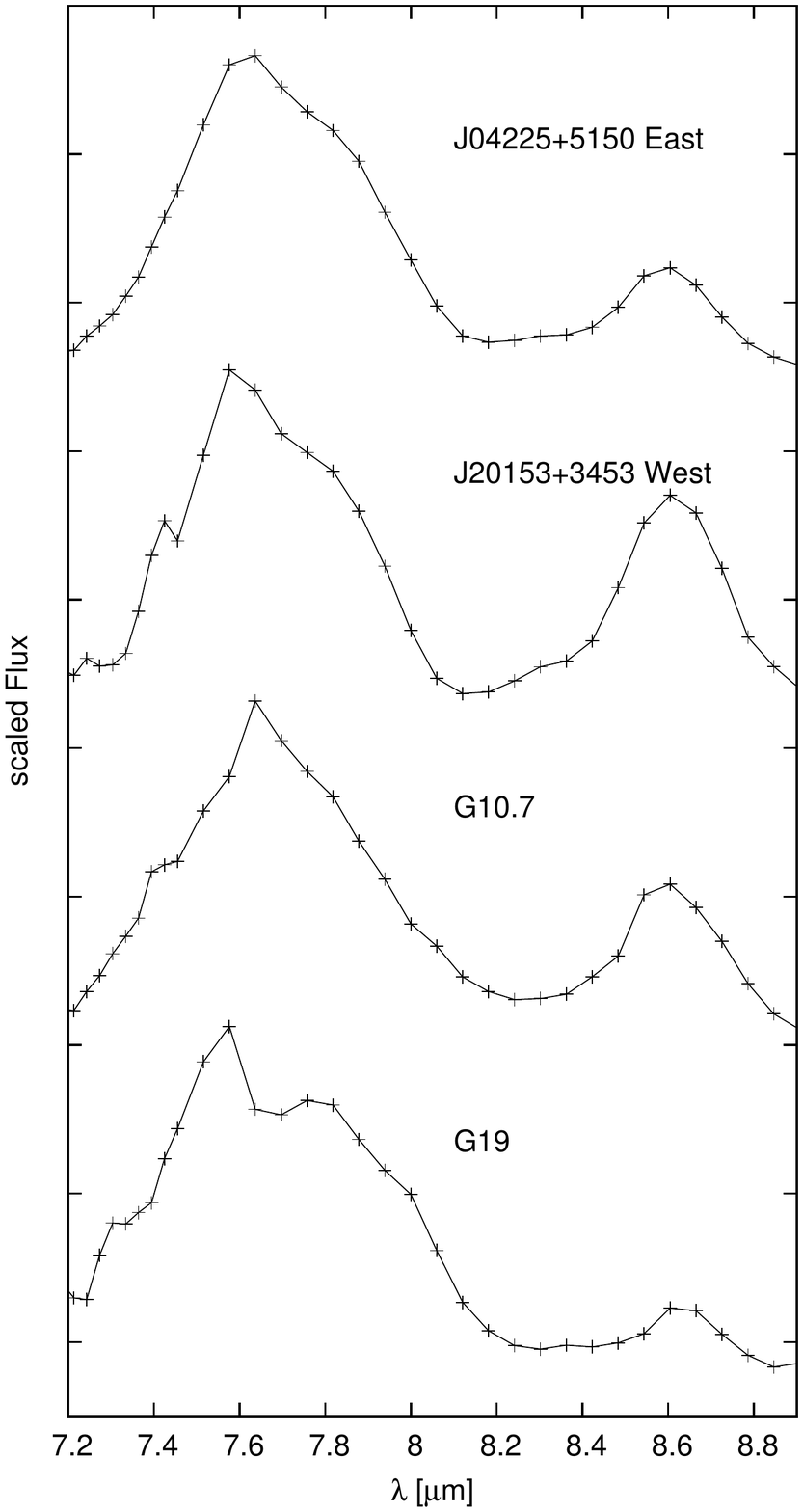}
 \end{minipage}
 \begin{minipage}[b]{.3\linewidth}
      \includegraphics[width=1\textwidth]{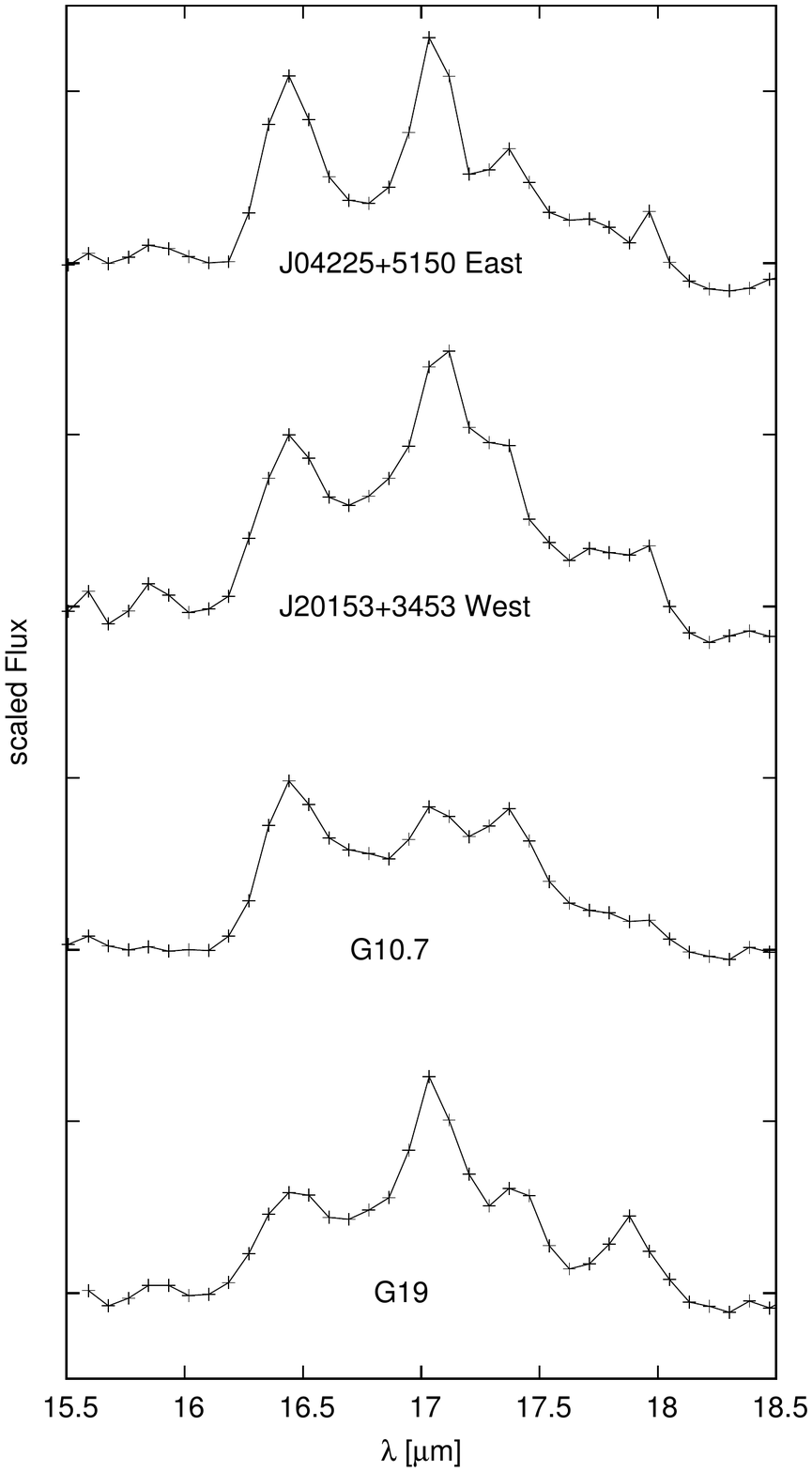}
 \end{minipage}
\caption{The 6.2~$\mu$m (left), 7.7/8.6~$\mu$m (center), and 16/18~$\mu$m (right) PAH bands. The global and local continuum has been removed, as proposed by \citet{Peeters02}.  The 17.03~$\mu$m feature is attributed to the H$_2$ $S(1)$ line. The PAH features for G19.27$+$0.07 were obtained from the background spectrum.}
\label{fig:PAH}
\end{figure*}

The PAH bands were identified with the features tabulated in \citet{Draine07}. 
The 6.2~$\mu$m and 7.7~$\mu$m features originate from pure CC stretching and CH in-plane bending. They, along with the 8.6~$\mu$m CH in-plane vibration mode \citep{Tielens08}, can be clearly detected in ISOSS~J04225+5150, ISOSS~J20153+3453 West, and G010.70-0.13 II (see Figure \ref{fig:PAH}). These bands are also observed in the background spectrum of G019.27+0.07, but not in the source spectrum. 
For these targets the first feature is centered at 6.21-6.22~$\mu$m and the PAH emission complex around 7.7~$\mu$m is asymmetrical with a peak between 7.57 and 7.64~$\mu$m. In each case, a secondary, weaker component is embedded in the red flank of this profile. Based on the profile shape and peak position, we conclude that class $\mathcal{A}$ PAHs are dominant in these regions \citep{Peeters02}. The 11.2~$\mu$m and 12.8~$\mu$m features can also be clearly identified in the spectra from these sources. These bands originate from CH out-of-plane bending modes \citep{Tielens08}. In addition, several PAH bands between 16.2 and 18.2~$\mu$m are present, although these features blend into a plateau (see Figure \ref{fig:PAH}, right panel). Similar plateaus have been observed with \textit{ISO} for some YSOs and compact \HII regions \citep{Kerckhoven00}. 
Another broad PAH band is clearly observed between $18.9$ and $19.0$~$\mu$m for ISOSS J04225+5150 East, ISOSS J20153+3453 West and G010.70-0.13~II. \citet{Tielens08} argued that this feature could arise from highly ionized PAHs.

\begin{figure*}[!htp]
	\centering
	\includegraphics[width=0.6\textwidth]{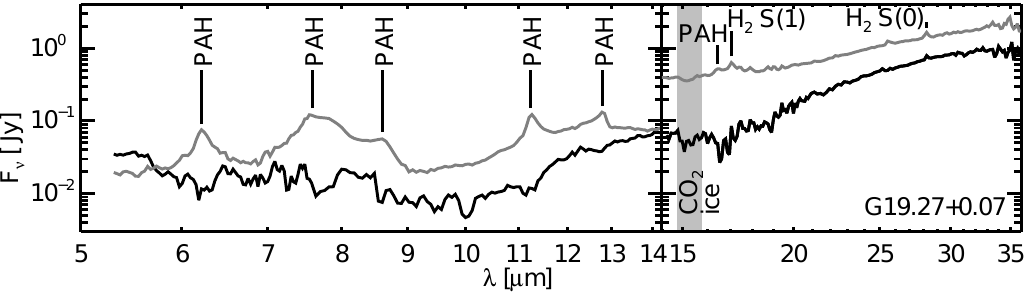}
	\caption{Spectra for G19.27-0.07 are shown. The source spectrum is extracted with a polynomial background fit and is plotted in black. The source background spectrum is plotted in gray.}
	\label{fig:spec_G19}
\end{figure*}

\begin{deluxetable*}{llllllll}
\tablecolumns{6}
\small
\tabletypesize{\footnotesize}
\tablewidth{0pt}
\tablecaption{Optical depths.}
\tablehead{\colhead{Source} 		& \colhead{$\tau_{H_2O}$} & \colhead{$\tau_{NH^+_4}$} & \colhead{$\tau_{CH_4}$} & \colhead{$\tau_{CO_2}$} &   \colhead{$\tau_{Silicate}\;^a$} & \colhead{A$_v\;^b$}}
\startdata
	ISOSS J04225+5150 East		& 						 &							&						  & $0.342 \pm 0.012$		& $0.57 \pm 0.03$				& 7.9 \\	
	ISOSS J18364$-$0221 West	& $0.256 \pm 0.017$		 &  $0.153 \pm 0.010$		& $0.055 \pm 0.009$ 	  & $0.51 \pm 0.04$			& $1.38 \pm 0.07$				& 19.2 \\
	ISOSS J23053$+$5953 East	& 						 &							&						  & $0.34 \pm 0.11$		    & $4.8 \pm 0.3$					& 67 \\
	G10.70$-$0.13 I				& $0.34 \pm 0.05$		 & $0.24 \pm 0.04 $			& 						  & $0.51\pm 0.06$		    & $1.60 \pm 0.08$				& 22 \\
	G10.70$-$0.13 II			& 						 &							&						  &							& $1.63 \pm 0.09 $				& 23 \\
	G025.04$-$0.02				& $0.49\pm 0.05 $		 & $0.399\pm  0.017$			& $0.118 \pm 0.013$	  &							& $ 2.5\pm 0.2$					& 35 \\
	G28.04$-$0.46				& $0.84 \pm 0.08$		 & $0.33\pm 0.04$			& 						  &						    & $1.01\pm 0.06$				& 14.1 \\ 
	G34.43$+$0.24				& $1.14 \pm 0.14$	 	 & $0.44 \pm 0.04$			& $0.28 \pm 0.05$		  &						    & $1.7 \pm 0.09$				& 24 \\
	G053.25$+$0.04 I			& $0.348\pm 0.027$		 & $0.174\pm 0.012$			& $0.081 \pm 0.010$		  & 						& $0.267 \pm 0.013$				& 3.7 \\
	G053.25$+$0.04 II			& $0.43 \pm 0.08$		 & $0.80\pm 0.08 $			& $0.18 \pm 0.04$		  &						    & $0.47 \pm 0.03$				& 6.6 
\enddata
    \label{tab:tau}
	\tablecomments{$^a$ To calculate the silicate optical depth $\tau_{9.7}$ after the correction for possible PAH contamination we used the \texttt{PAHFIT} routine \citep{Smith07}. $^b$ The optical extinction A$_v$ was calculated from the silicate bands using the extinction model from \citet{Robitaille07}.}
\end{deluxetable*}

The 11.2~$\mu$m band can appear as an extended bright background emission which is not associated with the source itself. This is the case for ISOSS J18364$-$0221, ISOSS~J23053+5953, G028.04-0.46, G034.43+0.24, and G053.25+0.04~I+II.
The same effect is also present for the 12.8~$\mu$m PAH band for G028.04-0.46 and G053.25+0.04~I.
No PAH bands were detected in the background subtracted spectrum from G010.70-0.13 I, although the 6.2~$\mu$m feature was observed as an extended background emission.
No PAH features were found in the spectrum from ISOSS J18364$-$0221 West.

\subsection{Silicates and molecular absorptions}

We were able to identify several absorption bands in our sample. The 9.7~$\mu$m silicate absorption feature was found in most sources of our sample. For ISOSS J20153+3453 West a broad emission component appears between 8 and 13~$\mu$m instead of the silicate absorption.
We  identified absorption bands by water ice (6.0~$\mu$m), NH$^+_4$ (6.85~$\mu$m), methane ice (7.67~$\mu$m) and carbon dioxide ice (15.2~$\mu$m). Those absorption bands and their optical depths are listed in Table~\ref{tab:tau}.
The IRDC sources, only taken in the short wavelength orders, are not accessible for CO$_2$ ice detections.
To calculate these optical depths, we fitted the spectral continuum by using multiple blackbody components and removing possible PAH contamination. This was accomplished with the \texttt{PAHFIT} routine described in \citet{Smith07}.
We calculated the optical depth $\tau$ for the ice features as in \citet{Quanz07} using
$$\tau =-\ln\left(\frac{F_{obs}}{F_{min}}\right)\;.$$ 
Here, $F_{min}(\lambda)$ is the observed flux of the absorption feature and  $F_{cont}(\lambda)$ is the fitted continuum flux. 
In addition to the removal of PAH contributions we utilized the \texttt{PAHFIT} routine to calculate $\tau_{9.7}$ for the silicate absorption. 
The routine is fitting the underlying continuum via multiple blackbody fits and using the silicate profile from \citet{Kemper04}. (For a detailed description see \citet{Smith07}.) The optical extinction $A_v$ calculated from $\tau_{9.7}$ is tabulated in Table \ref{tab:tau}.

\begin{figure*}
	\centering
	\plotone{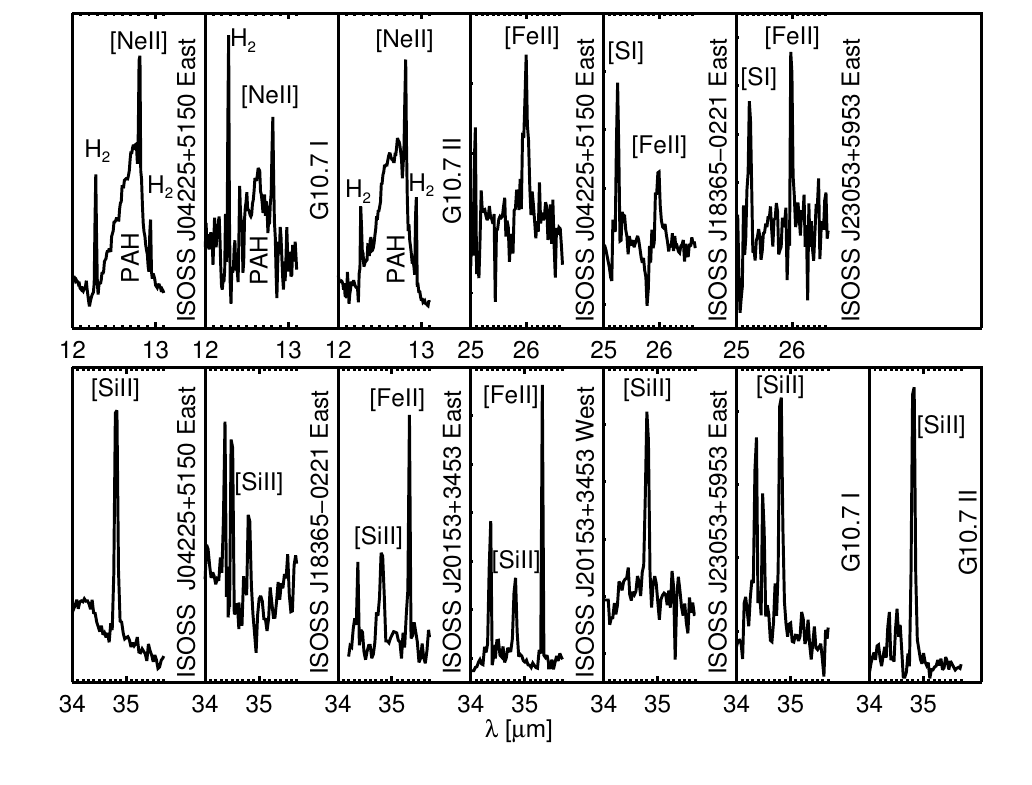}
	\caption{Fine structure lines as observed in the high-resolution spectra.}
	\label{fig:hispec}
\end{figure*}

\subsection{Ionic fine structure lines}
\begin{deluxetable}{lccccc}[tbh]
\tablecolumns{6}
\small
\tablewidth{0pt}
\tablecaption{Line flux ratios for \NeII}
\tablehead{
\colhead{Source} & 
\colhead{$\frac{F_{\mathrm{[Ne~II]}}}{ F_{\mathrm{S(2)}}}$}	& \colhead{$\frac{F_{\mathrm{[Ne~II]}}}{ F_{\mathrm{S(1)}}}$}	  & \colhead{$\frac{F_{\mathrm{[Ne~II]}}}{ F_{\mathrm{S(0)}}}$}	& \colhead{$\frac{F_{\mathrm{[Ne~II]}}}{ F_{\mathrm{[Fe~II]}}}$} & \colhead{$\frac{F_{\mathrm{[Ne~II]}}}{ F_{\mathrm{[Si~II]}}}$}	 }
\startdata 
ISOSS & & & & & \\
J04225$+$5150 			&	1.04	&	0.63 	& 0.64	& 0.22		& 0.04 \\
G010.70-0.13 I			&	2.8		&	0.36	& 0.04	& -			& 0.10 \\
G010.70-0.13 II			&	2.3		&	0.85	& 0.44	& 0.25		& 0.5 
\enddata
\label{tab:NeIIratios}    
\end{deluxetable}

Several fine structure lines were detected in the high resolution spectra. The identified lines are listed in Table \ref{tab:Lbol} and plotted  in Figure \ref{fig:hispec} after application of a second order baseline fit. The iron line \FeII can be detected at 25.99~$\mu$m or 35.35~$\mu$m. The \SiII appears at 34.82~$\mu$m. The sulfur line \SI (25.25~$\mu$m) is only detected for two sources in our sample. The neon emission \NeII  (12.81~$\mu$m) always appears on top of a broad PAH feature.

\subsection{Spatial extent of lines} \label{sec:res_spatial_extend}
The spatial line fits for the ISOSS J20153+3453, ISOSS J04225$+$5150 and the ISOSS J23053+5953 regions are shown in Figures \ref{fig:spatial_line_20} , \ref{fig:spatial_line_04}, and \ref{fig:spatial_line_23}, respectively. In these figures the relative line fluxes are shown as a function of their spatial distributions. The positions for the line extraction are overlaid for the short (SL) and long (LL) wavelength orders and the two different nod observations (in red and blue). For the LL slit positions we indicate only the overlap of the two nodded observations on the IRAC map and not the whole slit length. On top we present the spatial line fits for the long wavelength orders (14-37~$\mu$m) for the $S(1)$ 0-0 H$_2$ line and the 16.45~$\mu$m PAH feature. The lines were fitted with a second-order polynomial for local background estimation and a Gaussian for the line fit. The bottom plots present the line fits for short wavelength orders ($5.5-14.5$~$\mu$m) for PAHs and molecular hydrogen lines $S(5)$ and $S(3)$.

For the ISOSS J20153+3453 region we find similar spatial profiles of the 6.2 and 11.2~$\mu$m PAH features, which both peak slightly eastward ($\sim4-5''$) of the western MIPS maximum at 24~$\mu$m. 
The line fluxes from both these PAH features decrease more quickly toward the east than toward the west. The relative line flux distribution of these features is similar towards the western end of the slit, but, unlike the 11.2~$\mu$m feature, the 6.2~$\mu$m PAH band cannot be detected eastwards of the eastern MIPS source at 24~$\mu$m. The two molecular hydrogen lines $S(3)$ and $S(5)$ are not detectable at the western 24~$\mu$m MIPS peak location, but instead 
from two extended features to the east and west of this position. 

For ISOSS J23053+5953 East we detect an extended PAH feature at 11.2$\mu$m with almost constant line flux over the entire slit. Although the 6.2~$\mu$m PAH feature is not detected in the whole slit, it does have a line flux distribution similar  to the 11.2~$\mu$m feature in the western end. The S(5) H$_2$ feature is detected to both the east and west of the MIPS peak at 24~$\mu$m. The S(3) H$_2$ line is also detected at the MIPS peak position. 
The 16.45~$\mu$m PAH feature is only detected in the vicinity of the MIPS 24~$\mu$m source , whereas the $S(1)$ hydrogen line is extended over more than $180''$ in this region.

The spatial line fits for the ISOSS J04225$+$5150 region are shown in Figure \ref{fig:spatial_line_04}.
The line fluxes for the PAH features at 6.2~$\mu$m and 11.2~$\mu$m and the $S(3)$ H$_2$ line peak at the position of the brightest source observed in the IRAC and MIPS bands. The decrease of the line fluxes matches the morphology observed in the IRAC data. The slit for the LL observations partially covered the elongated submillimeter structure observed by SCUBA. While the 16.45~$\mu$m PAH and the H$_2$ $S(1)$ line fluxes decrease in the near vicinity of the central source, the H$_2$ line appears to be almost constant through the western submillimeter peak. 

\section{Discussion of individual sources} \label{sec:Discuss}
\notetoeditor{Figure \ref{fig:slit_20EW}, \ref{fig:slit_04E}, \ref{fig:slit_G10}, \ref{fig:slit_G10}, \ref{fig:slit_18E}, \ref{fig:slit_23E}, \ref{fig:slit_18W} should be fitted in a single column, placed directly under the corresponding caption. Figure \ref{fig:slit_IRDC} Figure should be fitted to the whole pagewidth. Figure \ref{fig:morph_20} and \ref{fig:CO2_18W} should be fitted into one column.}

All the sources exhibit a complex morphology in the infrared. Therefore, their spectra must be reviewed in the context of previous observations. The different sources will be discussed individually, sorted by their evolutionary stage. The first sources are the most evolved ones, with indications for PDRs, followed by young, deeply embedded sources with several absorption features.


\subsection{The ISOSS J20153$+$3453 region} \label{sec:20EW}
\begin{figure}[h]
 	\centering
	\plotone{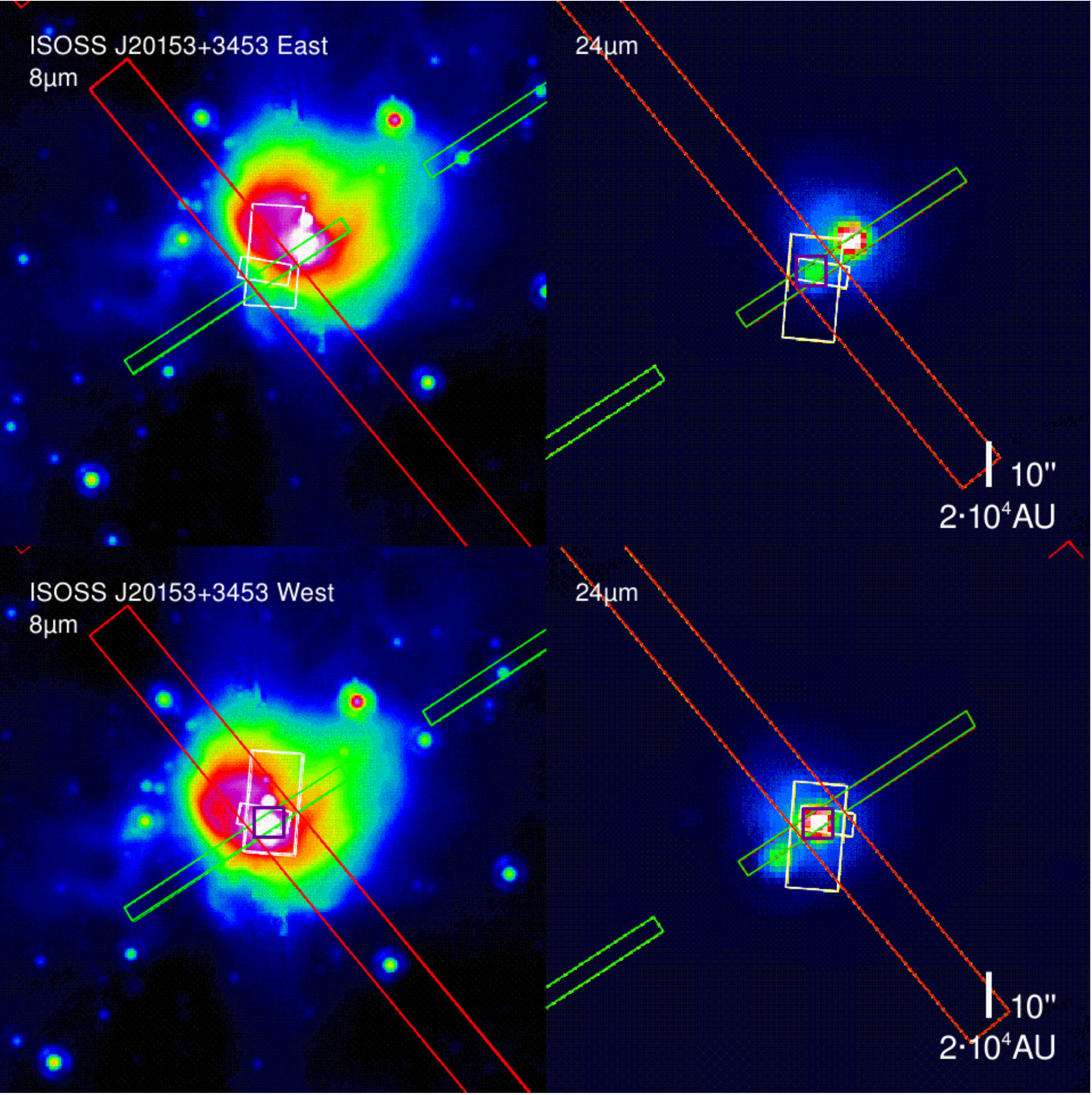}
	\caption{Overlay of the IRS slits for ISOSS
J20153$+$3453. The IRAC 8~$\mu$m (left) and MIPS 24~$\mu$m (right) images are shown in false colors. North is up and east is left. The slit positions for the IRS spectroscopy are overplotted for the low-resolution spectra in green (short wavelengths, SL) and red (long wavelengths, LL). The white overplots represent the high-resolution AOTs. The first nod position is indicated in the left panel and the second is in the right panel. The source is indicated by the dark magenta box.}
	\label{fig:slit_20EW}		
\end{figure}

\begin{figure}[!h]
 	\centering
	\plotone{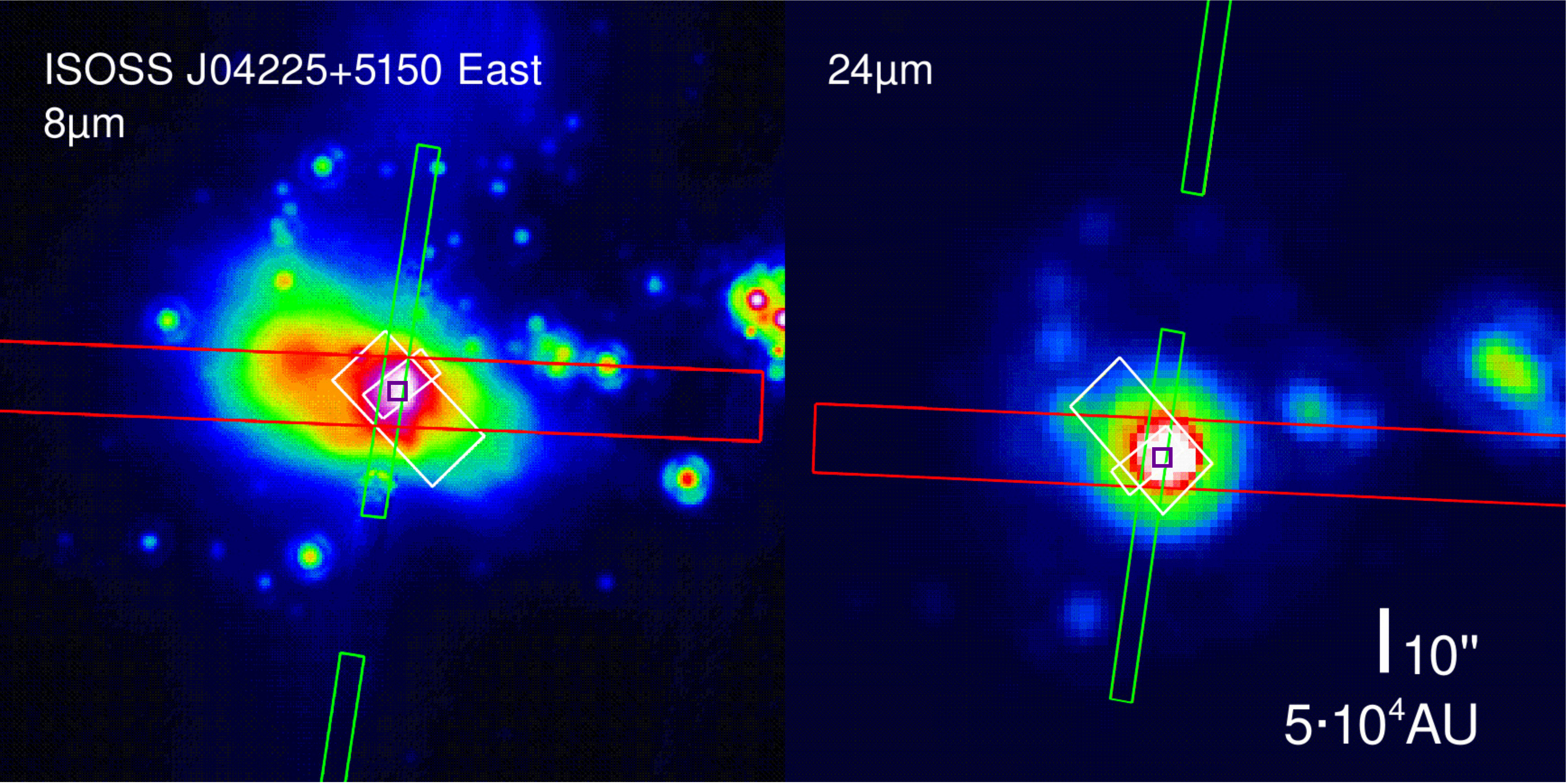}
	\caption{Overlay of the IRS slits for ISOSS~J04225$+$5150 East. Notations are the same as those for Figure~\ref{fig:slit_20EW}.	  
    }
	\label{fig:slit_04E}
\end{figure}

\begin{figure*}[ht]
 	\centering
	\plotone{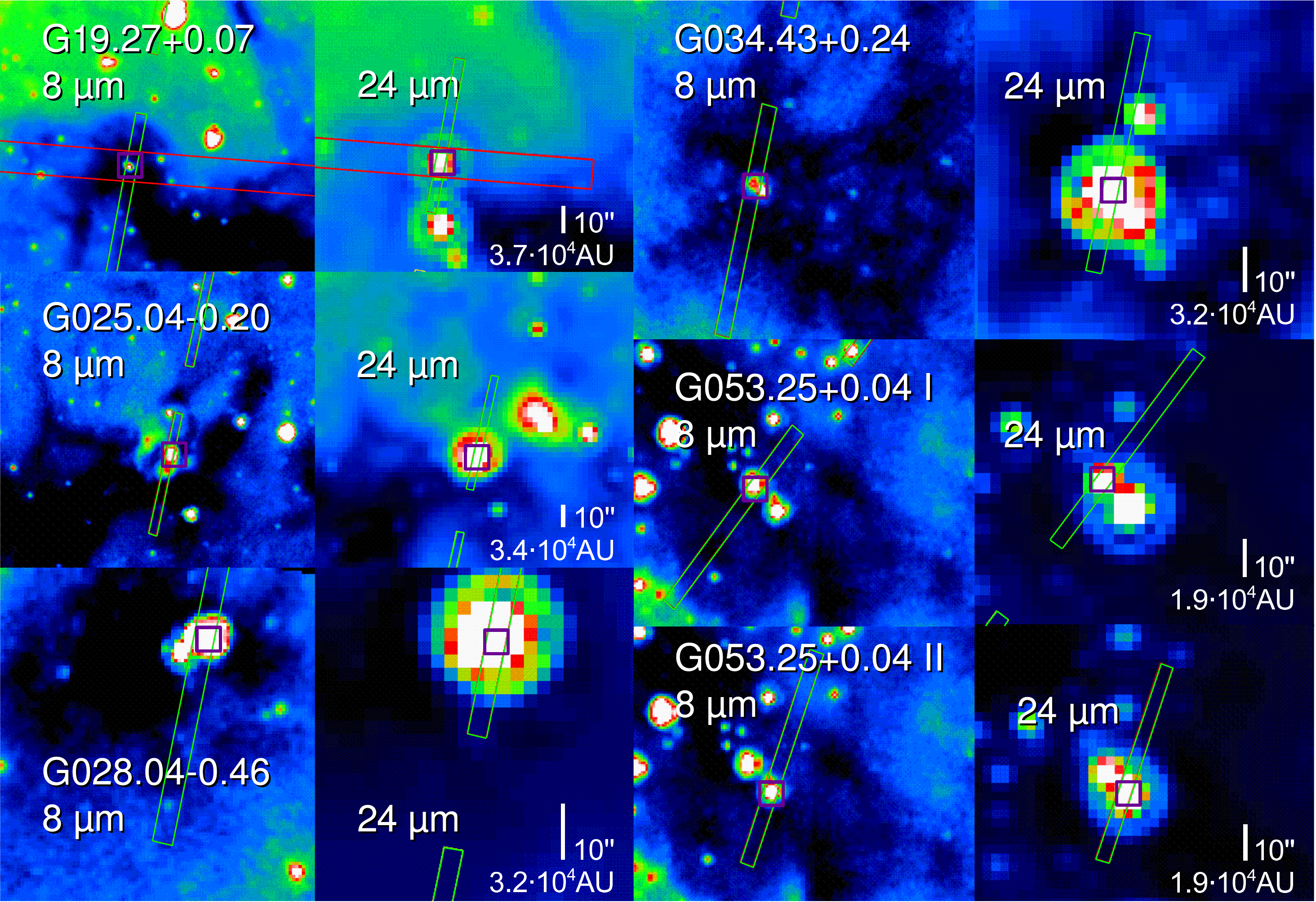}
	\caption{Overlay of the IRS slits for the IRDC sources. Notations are the same as those for
          Figure~\ref{fig:slit_20EW}.}
          	\label{fig:slit_IRDC}
\end{figure*}

\placefigure{fig:slit_20EW}

The ISOSS J20153$+$3453 region, located 2.0kpc away, was studied by \citet{Hennemann08_20}. In the MIPS
observations at 24~$\mu$m, the clump is resolved into a eastern and
western components with a separation of 10''. A single submillimeter
clump was detected close to the eastern source by SCUBA.  A dust
temperature of $T_d=15.3-17.0$~K and a gas mass of $M=87-149\;M_{\odot}$
were calculated for the clump. Based on the IRAC colors, the MIPS
source at 24~$\mu$m associated with the submillimeter peak was assumed to be a class 0/I object. 
The mass reservoir in this region contains enough gas to form a massive star, but due to the multiplicity of NIR and IRAC sources the interpretation of the further evolution in this region is not straightforward. 
Another MIPS 24~$\mu$m source, surrounded by extended emissions, is located on the northwestern limb of the submillimeter clump.

\begin{figure}
 	\centering
 	\plotone{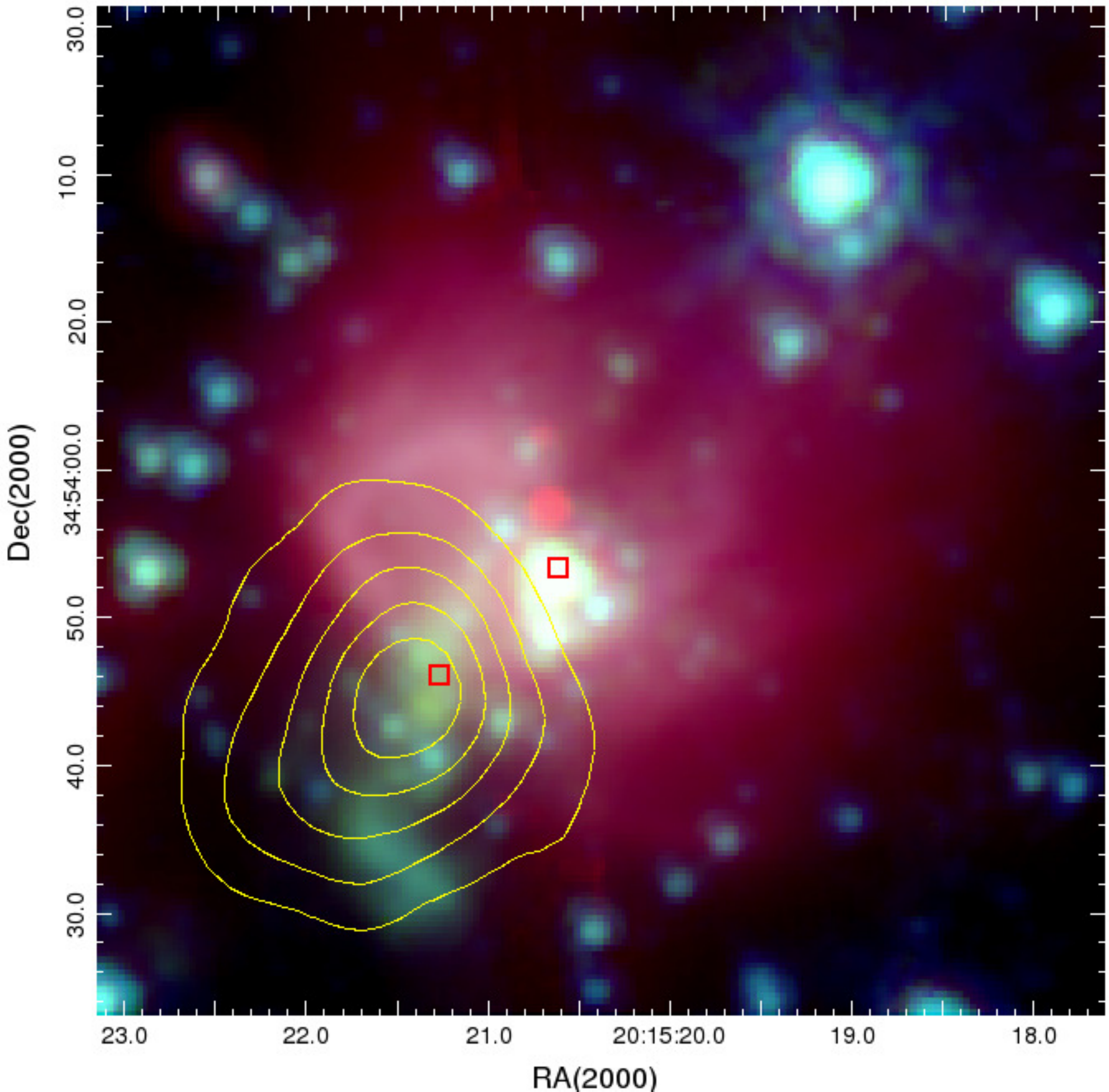}
 	\caption{IRAC composite image of the ISOSS J20153$+$3453 region: the 3.6~$\mu$m channel is colorized in blue, 4.5~$\mu$m in green, and 8.0~$\mu$m in red. The contours of the SCUBA observations at 450~$\mu$m are shown in yellow. The two MIPS point sources at 24~$\mu$m are indicated by the red box points. They are used for the IRS pointing. At the position of the eastern MIPS sources and the SCUBA source an elongated ``green and fuzzy'' feature can be seen.}
 \label{fig:map_20h}
\end{figure}

\begin{figure*}[htp]
 	\centering
	\includegraphics[width=0.63\textwidth]{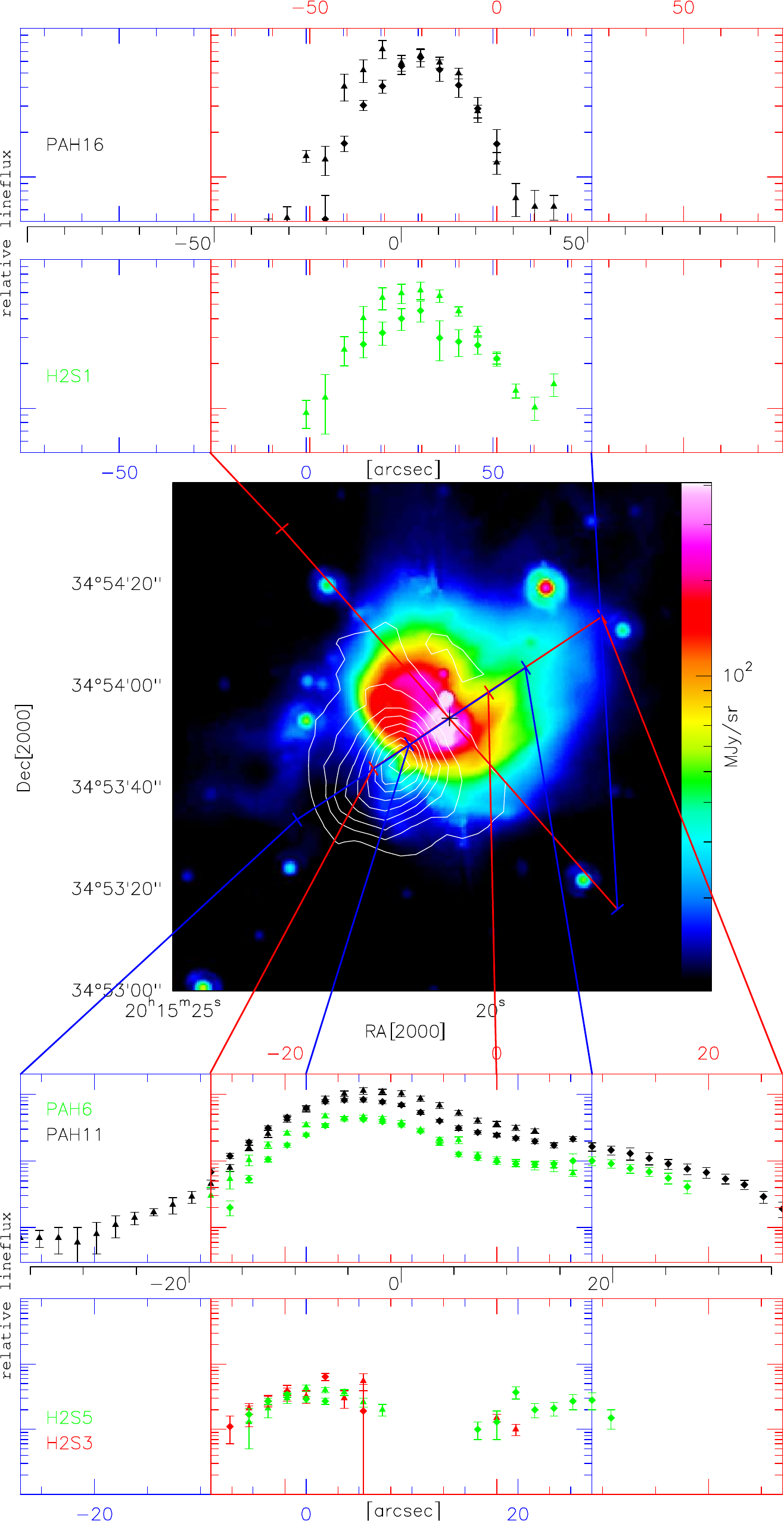}
	\caption{For the low resolution spectra the spatial evolution of several PAH and H$_2$ lines over the whole slit for the ISOSS~J20153+3453 region is presented. The IRAC 8~$\mu$m map is shown and the 450~$\mu$m SCUBA contours are overplotted. The slit overlay is explained in Section \ref{sec:res_spatial_extend}. }
	\label{fig:spatial_line_20}
\end{figure*}

\begin{figure*}[htp]
 	\centering
	\includegraphics[width=0.63\textwidth]{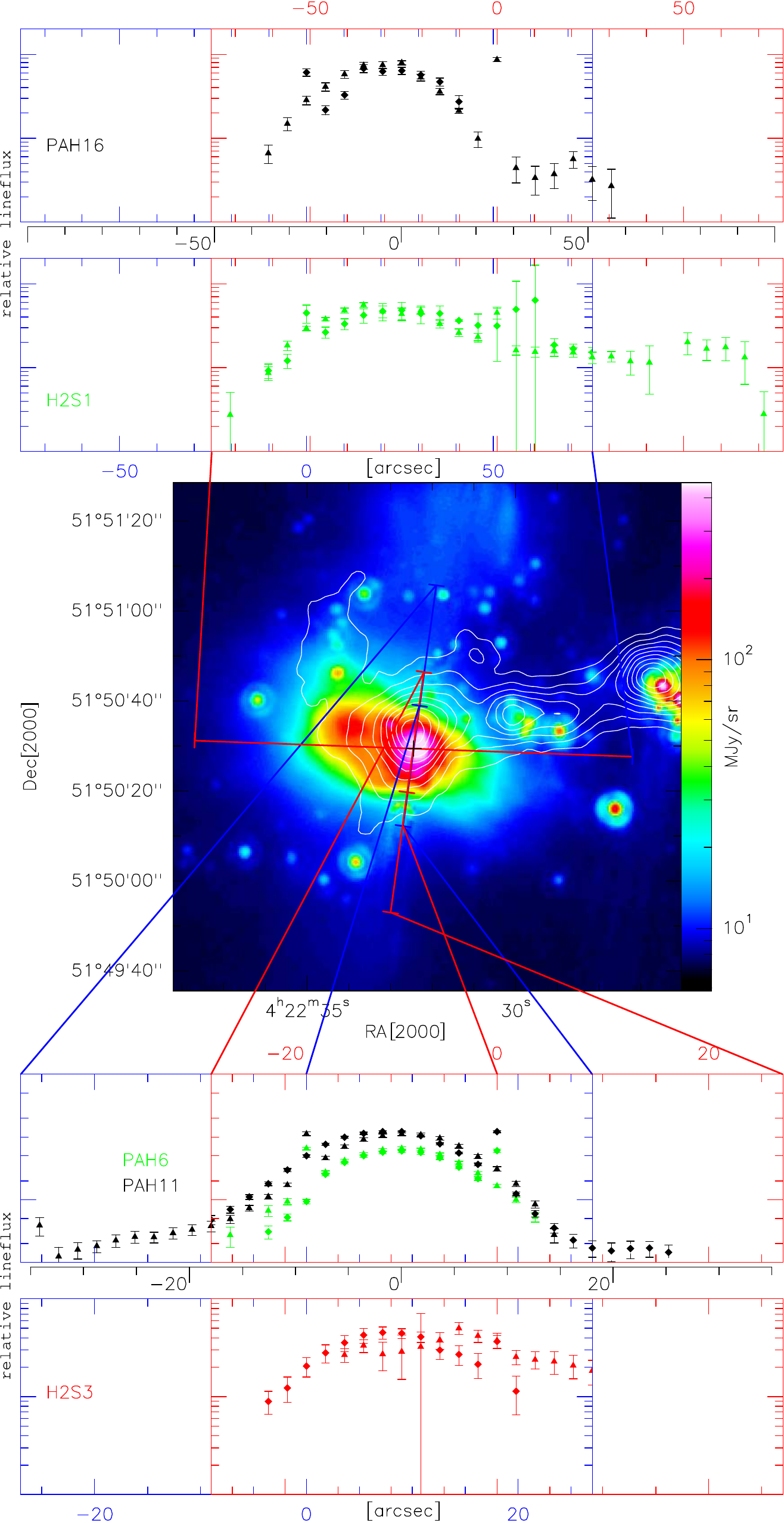}
	\caption{Spatial line fits as in Figure \ref{fig:spatial_line_20} for the ISOSS J04225+5150 East region. The IRAC image at 8~$\mu$m and the SCUBA contours at 450~$\mu$m are used.}
	\label{fig:spatial_line_04}
\end{figure*}

The IRS spectrum of the western MIPS  24~$\mu$m  source is dominated by several PAH bands.
The broad emission feature between 8 and 13~$\mu$m is
typical for small, transiently heated dust grains distributed over a wide spatial
range around the central sources.  Photoelectric heating of the small dust and PAH grains
accounts for thermal coupling of the grains to the gas, hence a
certain amount of the observed molecular hydrogen must be thermally
excited. The smooth spatial distribution of the $S(1)$ H$_2$ line
represents such thermally excited gas (Figure \ref{fig:spatial_line_20}).

As for the ISOSS~J23053$+$5953 region, the detected \SiII and \FeII lines and the spatial distribution of the $S(5)$ and $S(3)$ H$_2$ emission can be attributed to J-shock excitation. The elongated ``green and fuzzy'' structures on the IRAC composite images fit this picture as well (see Figure \ref{fig:map_20h}).
In contrast to ISOSS~J23053$+$5953, the PAH features in the spectrum of ISOSS~J20153$+$3453 West are identified as class $\mathcal{A}$ according to \citet{Peeters02} (see Section \ref{Intro}). Therefore, UV field strengths are sufficient to form a PDR.
The 18.9~$\mu$m feature could represent highly ionized cationic PAHs \citep{Tielens08} and give further evidence for a strong UV radiation. The detected \SiII and \FeII lines are found not only for J-shocks; they are known as minor coolants in PDRs \citep{Abel05,Hollenbach97, Kaufman06}. We tried to fit the observed H$_2$ line ratios to the PDR models by \citet{Kaufman06} (see Figure \ref{fig:PDRmodel}). The contribution by shock excited H$_2$ emission might impair the results. Nevertheless, the models show a trend toward higher gas densities in the outer layers of PDR ($n_{\mathrm{H}}>10^5\;\textrm{cm}^{-3}$) and UV radiation fluxes between $10^2$ and $10^3\;\textrm{G}_0$. Similar results are found for ISOSS~J04225$+$5150 East and G010.70-0.13 (Sections \ref{sec:04E} and \ref{sec:G10}).
Although these observations can be interpreted as the result of a PDR, no indication of an associated \HII region can be found in the 20~cm NVSS data. Unfortunately, there are no 6~cm survey products available that are sensitive enough for the detection of an \HII region.

\begin{figure}
 	\centering
	\plotone{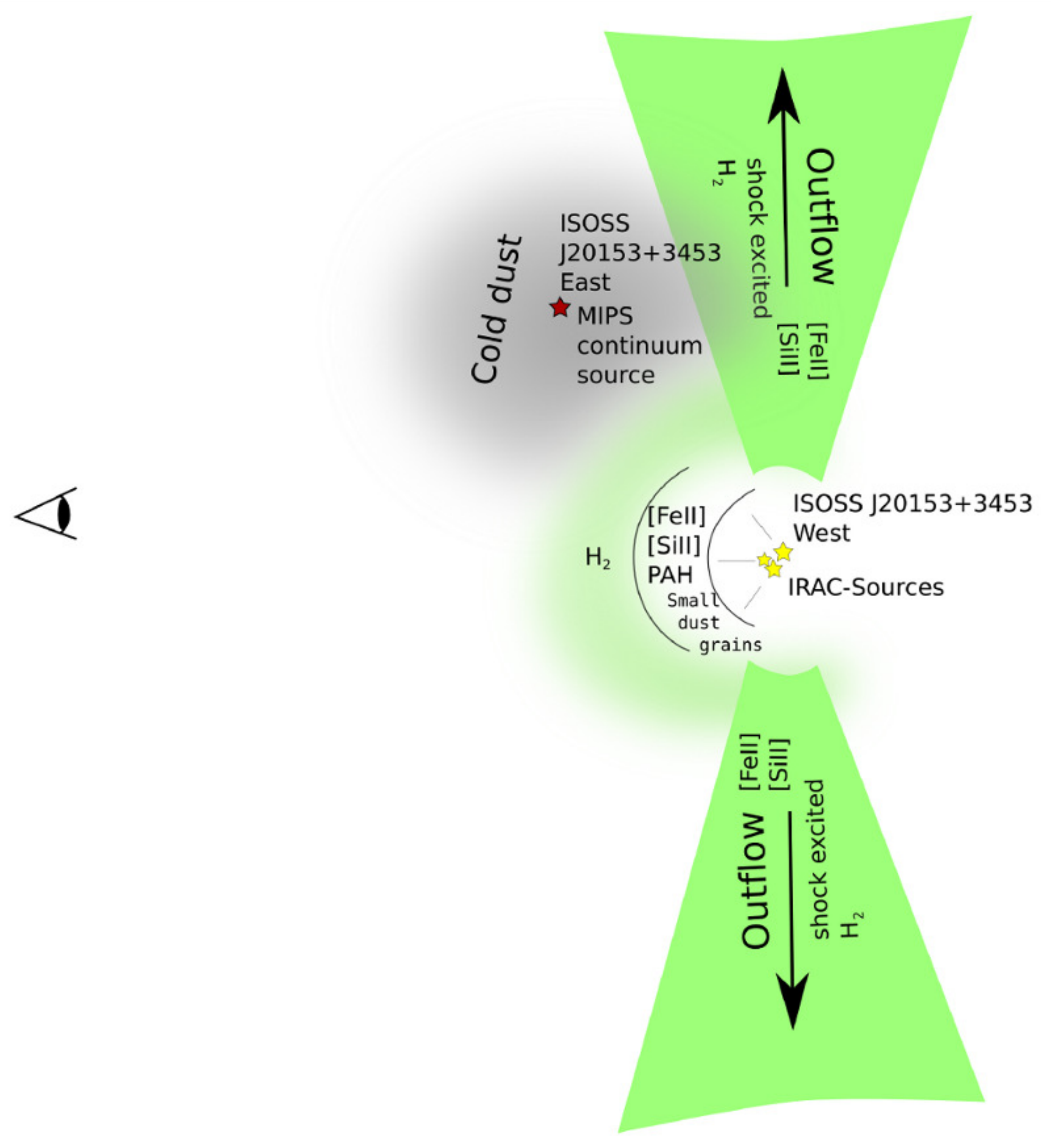}
	\caption{Schematic representation of the ISOSS J20153$+$3453 region.}
	\label{fig:morph_20}
\end{figure}

One possible interpretation of the morphology of this region is shown in Figure \ref{fig:morph_20}. We see the cationic PAHs and small grains photoelectrically heated by the IRAC sources at the western position. H$_2$ emission and atomic fine structure lines originate from the warm layers of a PDR, but have also been formed via shocks. A clump of cold dust is located eastward in the foreground, observed in the submillimeter regime (Figure \ref{fig:map_20h}). The LL spectrum extracted at the eastern position is centered on the eastern MIPS 24~$\mu$m peak. Here a young, deeply embedded continuum source is driving the heating of the submillimeter clump.


\subsection{ISOSS J04225$+$5150 East} \label{sec:04E}

\begin{figure}[h]
 	\centering
	\plotone{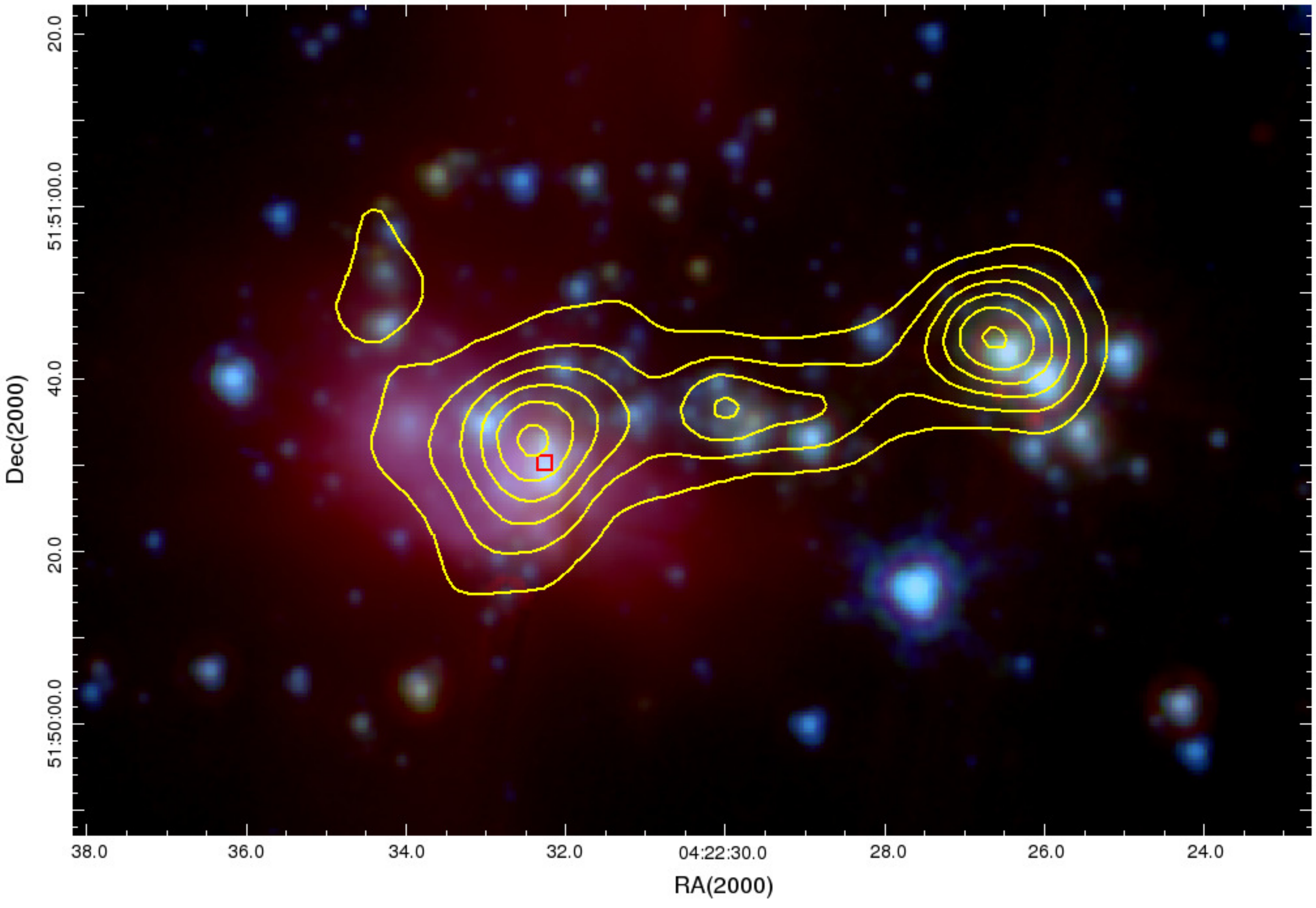}
 	\caption{IRAC composite image of the ISOSS J04225+5150 region: the 3.6~$\mu$m channel is colorized in blue, 4.5~$\mu$m in green and 8.0~$\mu$m in red. The contours of the SCUBA observations at 450~$\mu$m are overplotted in yellow. The peak position of the MIPS source at 24~$\mu$m , which is used for the IRS pointing, is indicated by the red box point.}
 \label{fig:map_04h}
\end{figure}

ISOSS J04225$+$5150 East shows three compact sources in the mid-infrared and submillimeter regime. The eastern clump has a mass of $M=510\;M_{\sun}$ and a temperature of $T_d=17$~K \citep{BirkmannPhD}.\\
The IRAC images show a bright point source with extended PAH emissions in the 8~$\mu$m channel, and diffuse H$_2$ emissions at 4.5~$\mu$m. At 24~$\mu$m, the source is surrounded by warm dust. Slightly shifted northward from the MIPS 24~$\mu$m peak position the center of a submillimeter dust clump is detected by SCUBA (Figure \ref{fig:map_04h}).
ISOSS J04225+5150 East shares several other features with the ISOSS~J20153$+$3453 region: a class $\mathcal{A}$ PAH spectrum, \SiII and \FeII lines. Hence, J-shocks and a PDR have been considered as the origin of these features as for ISOSS~J20153$+$3453 West.
However, in contrast to  ISOSS J20153$+$3453, we detected the presence of the \NeII line. \NeII typically appears towards \HII regions \citep{Abel05}. The presence of the \NeII line can also be explained by shock excitation. \citet{Hollenbach89} predicted the presence of a strong \NeII line only for J-shocks with high shock velocities. The \NeII line ratios in Table \ref{tab:NeIIratios} agree with the predictions from \cite{Hollenbach89}. 
The H$_2$ line ratios are similar to ISOSS J20153$+$3453 West, therefore the results are similar for the PDR modeling (Figure \ref{fig:PDRmodel}). The line ratio fitting favor models with higher gas density ($n_\mathrm{H}>10^5\textrm{cm}^{-3}$) and UV radiation fluxes between $10^2$ and $10^3\;\textrm{G}_0$.
For ISOSS J04225+5150, both PAH and H$_2$ are present, with almost constant line strengths in the western direction. The source is associated with an elongated structure of higher dust density observed with SCUBA at 450~$\mu$m.
The absorption feature at 10~$\mu$m originates from an outer ridge of dust.

\subsection{G010.70-0.13} \label{sec:G10}
\begin{figure}[h]
 	\centering
	\plotone{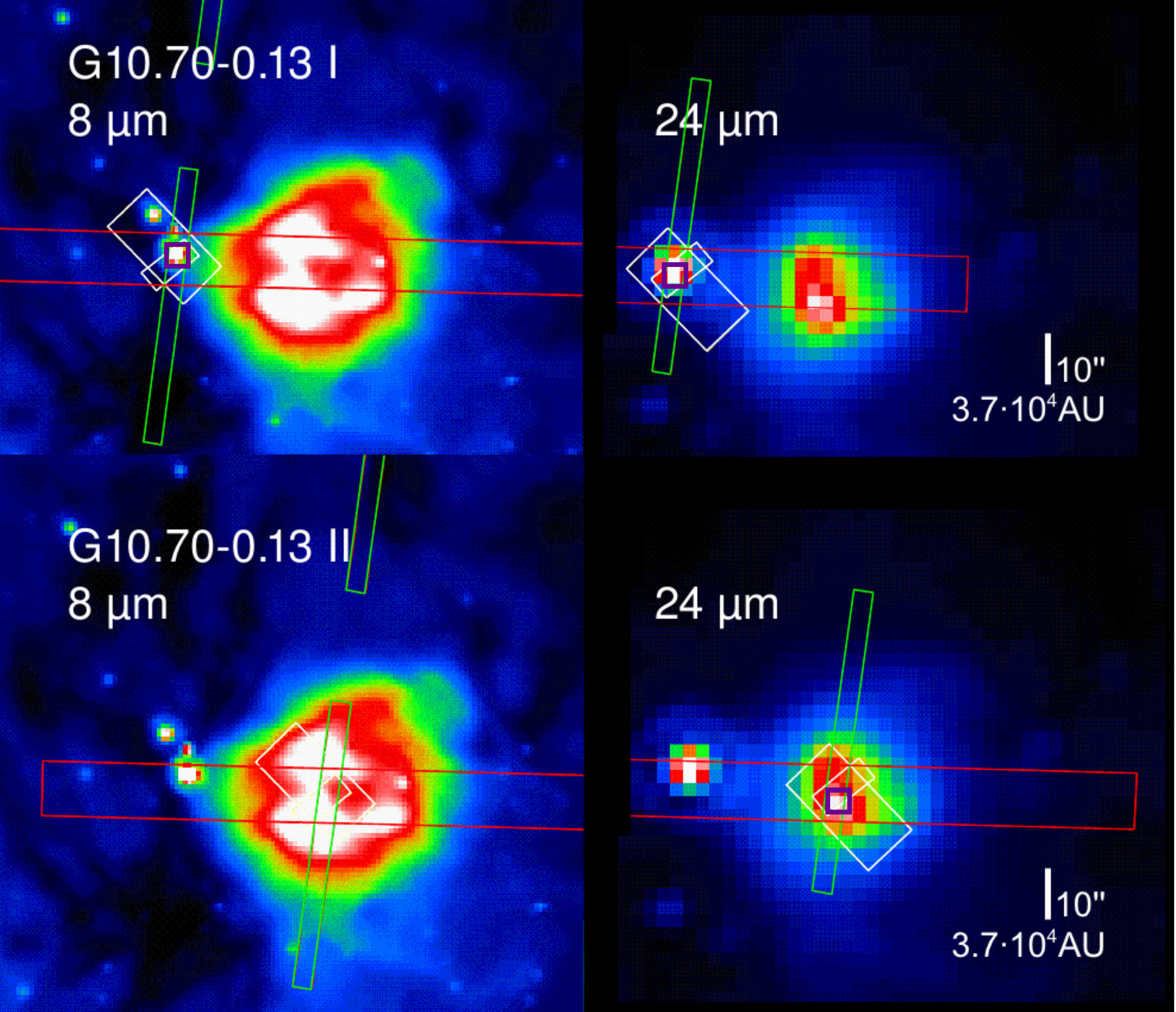}
	\caption{Overlay of the IRS slits for the G010.70-0.13 region. Notations are the same as those for
          Figure~\ref{fig:slit_20EW}.}
	\label{fig:slit_G10}
\end{figure}

\placefigure{fig:slit_G10}[!h]

\begin{figure}[b]
 	\centering
	\plotone{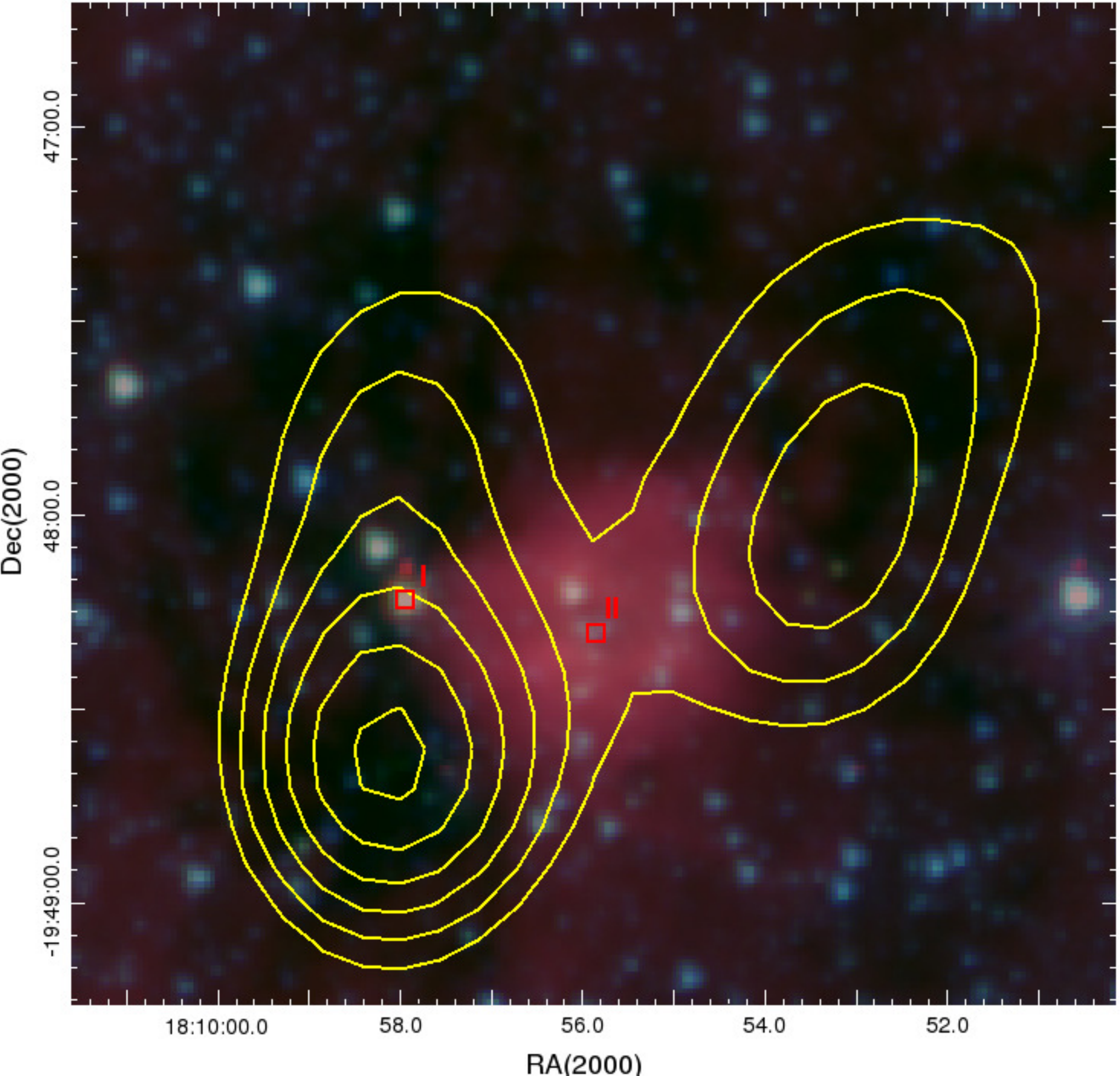}	
	\caption{IRAC composite image of the G010.70-0.13 region. The 3.6~$\mu$m channel is colorized in blue, 4.5~$\mu$m in green and 8.0~$\mu$m in red. The peak positions of the MIPS point sources at 24~$\mu$m are indicated by the red box points. The IRS slits were centered on the position.}
 \label{fig:map_G10.7}
\end{figure}

The G010.70-0.13 region can be found in the \textit{Spitzer} dark cloud catalog \citep{Peretto09}.
Two spectra were obtained for this region. The IRAC composite image shows dark filamentary structures against the bright PAH background. Around the eastern MIPS 24~$\mu$m peak position (G010.70-0.13~II), extended PAH emissions are present in the 8~$\mu$m band, as are warm dust emission in the MIPS maps at 24~$\mu$m. 
A second western MIPS 24~$\mu$m source is centered on a strong IRAC point source (G010.70-0.13~I). It is located in the tail of an extended submillimeter clump obtained from the SCUBA legacy program (see Figure \ref{fig:map_G10.7}).
For G010.70-0.13~I, we identified several absorption features, such as NH$^+_4$, water ice, and CO$_2$ ice, that are typical for deeply embedded sources. These are also observed for ISOSS~J18364$-$0221 West as well as the sources in our sample associated with young, embedded sources (discussed in section \ref{sec:Ygalactic} ). 
Even though no extended H$_2$ emissions were found, the flux ratios for \NeII, \SiII and \FeII (Table \ref{tab:NeIIratios}) for  G010.70-0.13 I+II agree with ratios for J-shocks \citep{Lahuis10}. 
The spectrum of G010.70-0.13~II, located to the west, is dominated by PAH bands. Bright UV radiation entails a class $\mathcal{A}$ PAH spectrum.

Similar to ISOSS~J04225$+$5150 East and ISOSS~20153$+$3453 this region shows indications for J-shocks and a PDR as well.
The comparison of the H$_2$ line ratios with the PDR in Figure \ref{fig:PDRmodel} shows a correlation for a gas density of $n \geq 10^5\;\textrm{cm}^{-3}$ and a UV flux of $1-2\times 10^2\;\textrm{G}_0$ for G010.7$-$0.13~II. The line ratio fit for G010.7$-$0.13~I  did not converge as for the prior spectrum; the eastern spectrum is likely just tracing the outer parts of the PDR, and the contribution of shocks to the H$_2$ lines is stronger in this spectrum.
We obtained the 6~cm and 20~cm survey products for this region from the Multi-Array Galactic Plane Imaging Survey \citep{Helfand06,White05}. No indications for an associated \HII region were found. Our hypothesis is that G010.70-0.13 contains a PDR connected to a very young \HII region, which is still optically thick at 6~cm and 20~cm. In such a case, the \HII region would not be detected given the sensitivity of these radio observations. 


\subsection{ISOSS J18364$-$0221 East} \label{sec:18E}
\begin{figure}[h]
 	\centering
	\plotone{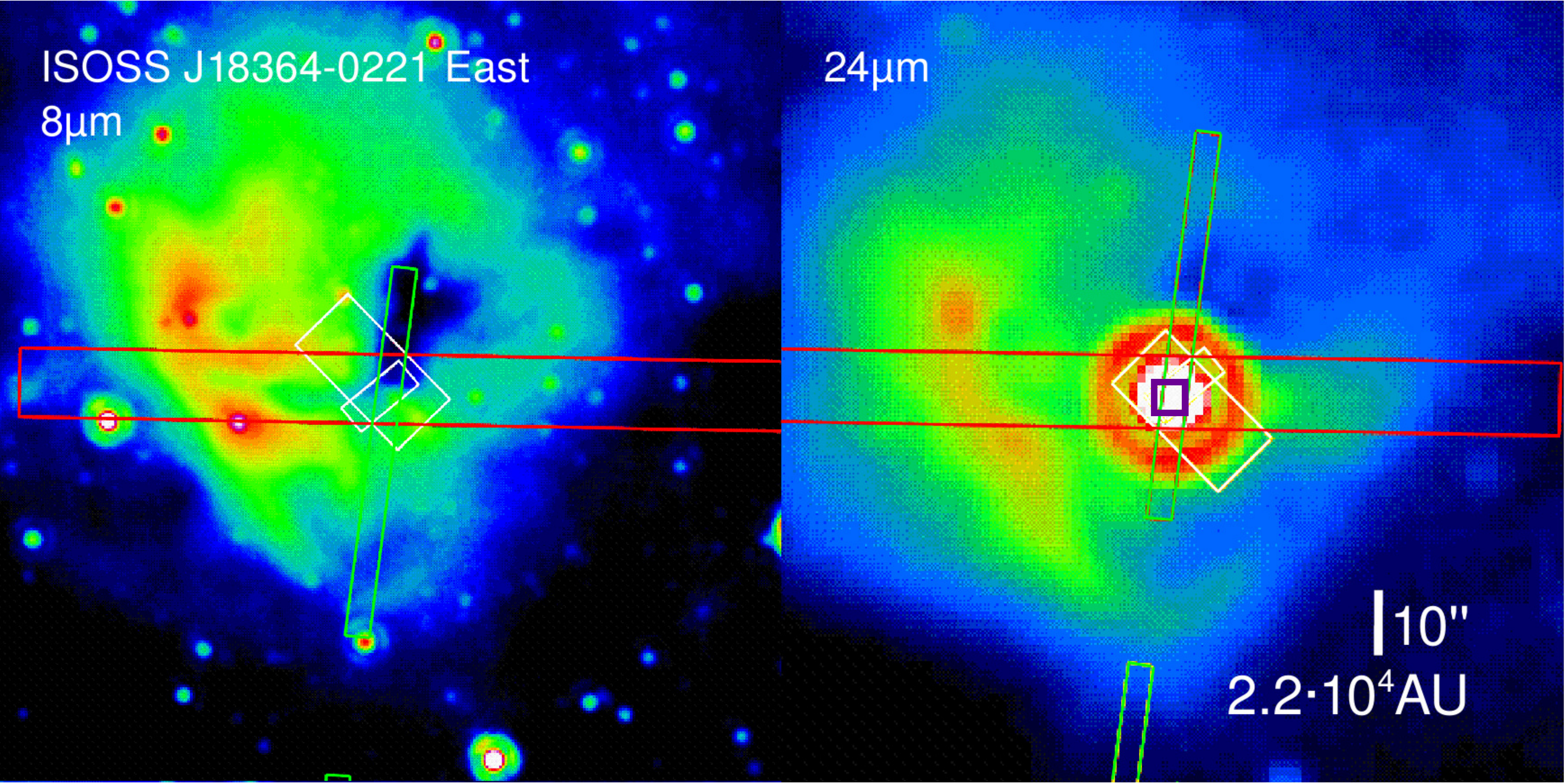}
	\caption{Overlay of the IRS slits for ISOSS
          J18364$-$0221 East. Notations are the same as those for
          Figure~\ref{fig:slit_20EW}.}
	\label{fig:slit_18E}
\end{figure}

\placefigure{fig:slit_18E}

Interferometric maps of 1.3 and 3.4~mm continuum emission show two cores. One core appears as a point source on the MIPS maps, surrounded by warm dust emissions at 24~$\mu$m. The IRAC observations do not show any point source at this position. The dust mass of both cores was estimated with $12\;M_\odot$ and $18\;M_\odot$, with a dust temperature of 22~K and 15~K.
Narrowband NIR observations at 2.122~$\mu$m (H$_2$ $S(1)$ 1-0) are related to outflow lobes, traced by CO(2-1) emissions around the MIPS source \citep{Hennemann09}.

The IRAC bands for 4.5~$\mu$m and 5.8~$\mu$m show some filaments, tracing H$_2$ stemming from the outflow. 
The MIPS 24~$\mu$m source is surrounded by extended emissions in the 8~$\mu$m IRAC band. This band reveals a brighter eastward ridge.

Although the SL orders only show a non-uniform extended background spectrum, a strong PAH band appears over the whole slit at 11.2~$\mu$m. The only detected lines are $S(1)$ H$_2$, \SI, \FeII and \SiII, which may be shock related. The \SI emission is not strong in PDRs \citep{Kaufman06}, but can be found, velocity independent, in J-shocks \citep{Hollenbach89}.
 
Due to the lack of the SL order spectra, no further conclusion can be drawn.


\subsection{ISOSS J23053$+$5953 East} \label{sec:23E}
\begin{figure}[h]
 	\centering
	\plotone{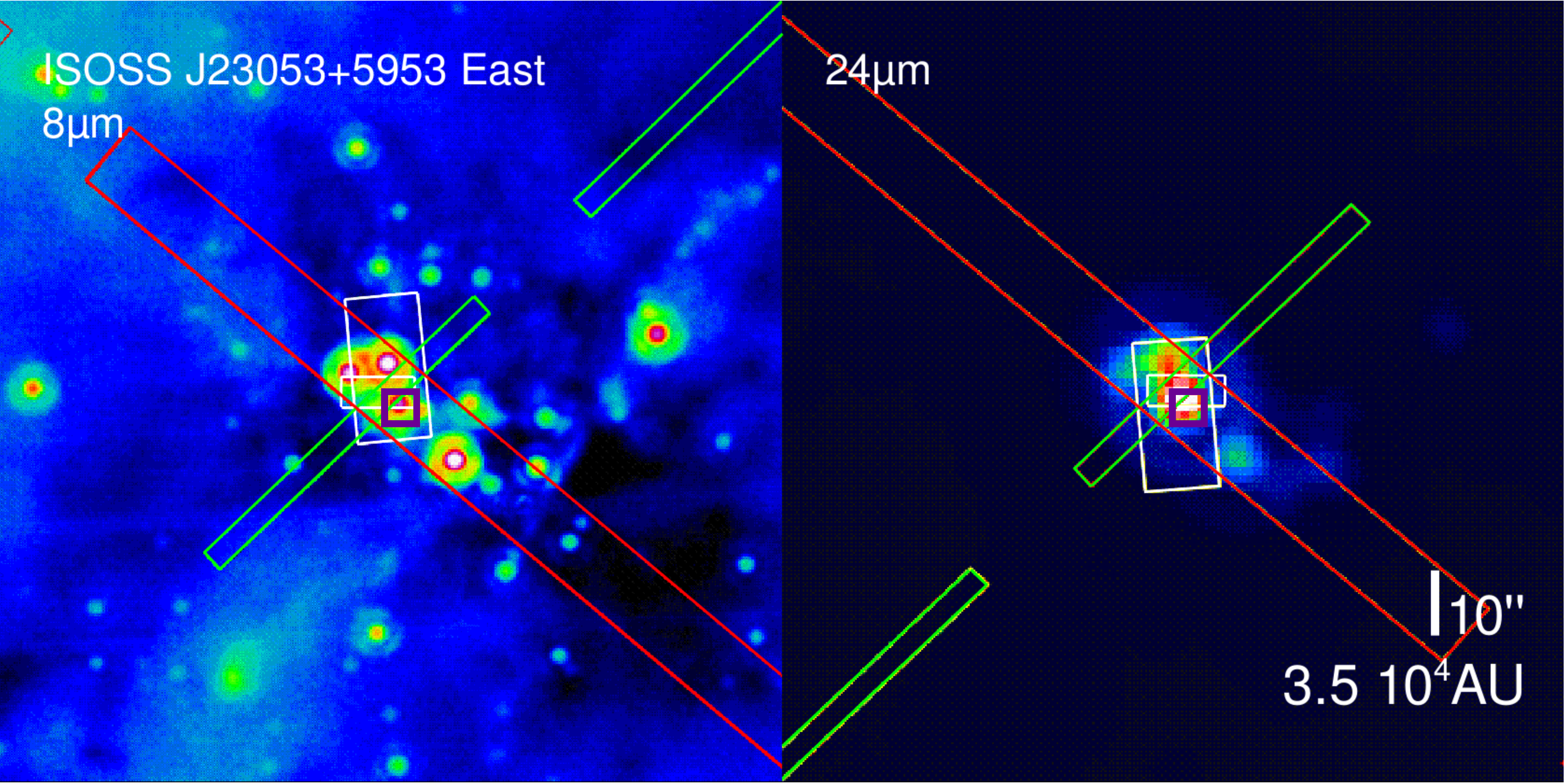}
	\caption{Overlay of the IRS slits for ISOSS J23053$+$5953
          East. Notations are the same as those for
          Figure~\ref{fig:slit_20EW}.}
	\label{fig:slit_23E}
\end{figure}

\placefigure{fig:slit_23E}

The ISOSS J23053$+$5953 region has been studied by \citet{Birkmann07}. It is located at a distance of 3.5kpc.  In this
region, two submillimeter cores with $T_d = 19.5$~K~/$17.3$~K and a mass of $\approx200M_{\sun}$ each were observed. Both cores show signs of infall, as indicated by interferometric HCO$^+$ measurements.  IRAC and NIR imaging show several embedded point sources, surrounded by diffuse PAH emission in the IRAC 4.5~$\mu$m
images. Narrow-band observations ($\lambda=2.122$~$\mu$m) taken by the Omega~2000 prime focus camera at the Calar Alto 3.5~m telescope show elongated H$_2$ $S(1)$ 1-0 shock excitation. 

The 24~$\mu$m MIPS observations show a bright source centered on the eastern submillimeter core, surrounded by extended warm dust emission and two fainter sources. Another faint source appears eastward, close to the peak position of the western submillimeter core.

\begin{figure*}[htp]
 	\centering
	\includegraphics[width=0.63\textwidth]{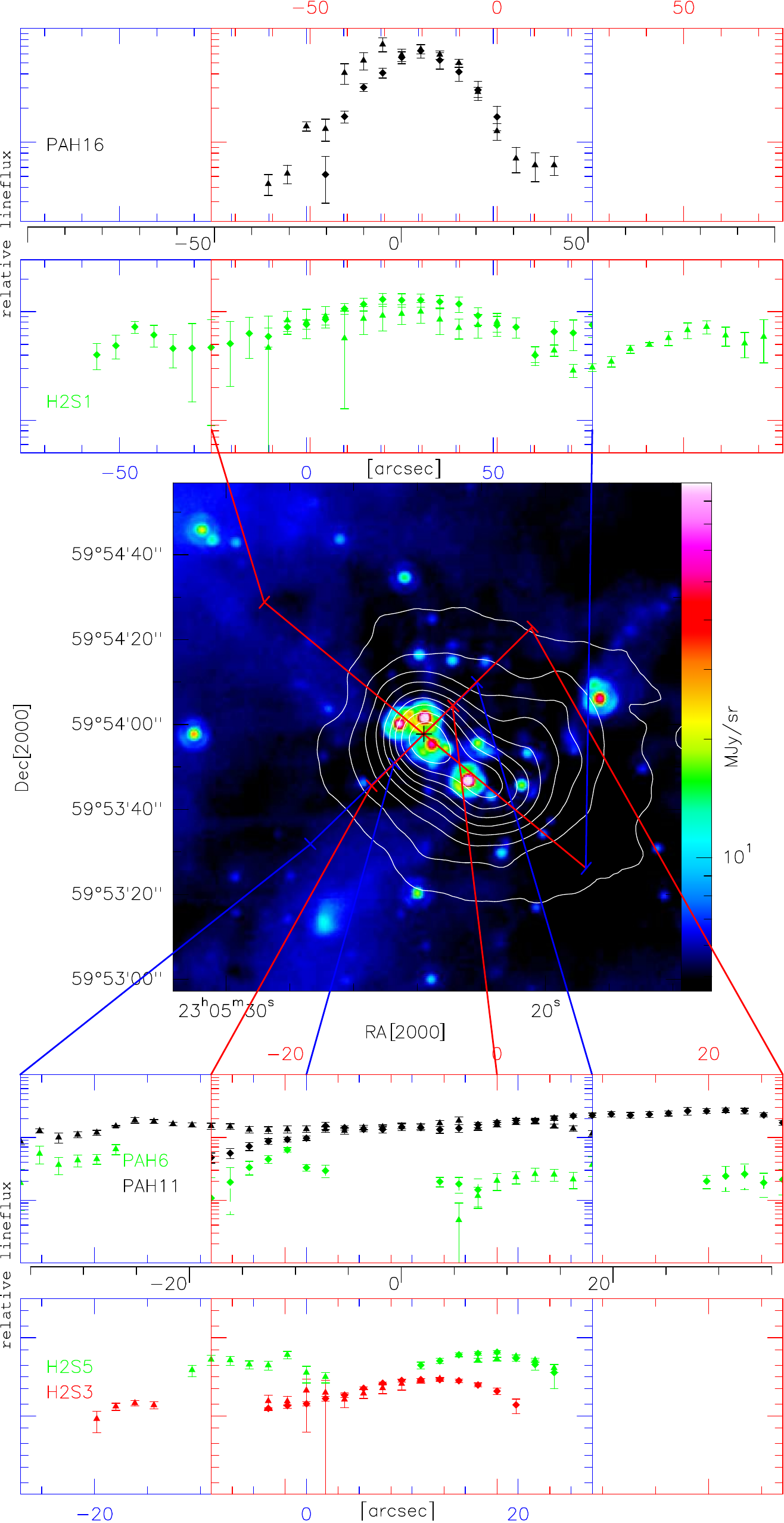}
	\caption{Spatial line fits are the same as in Figure \ref{fig:spatial_line_20} for the ISOSS J23053+5953 region. The IRAC image at 8~$\mu$m and the SCUBA contours at 450~$\mu$m are used.}
	\label{fig:spatial_line_23}
\end{figure*}

\begin{figure*}[hbp]
	\centering
	\includegraphics[width=.7\textwidth]{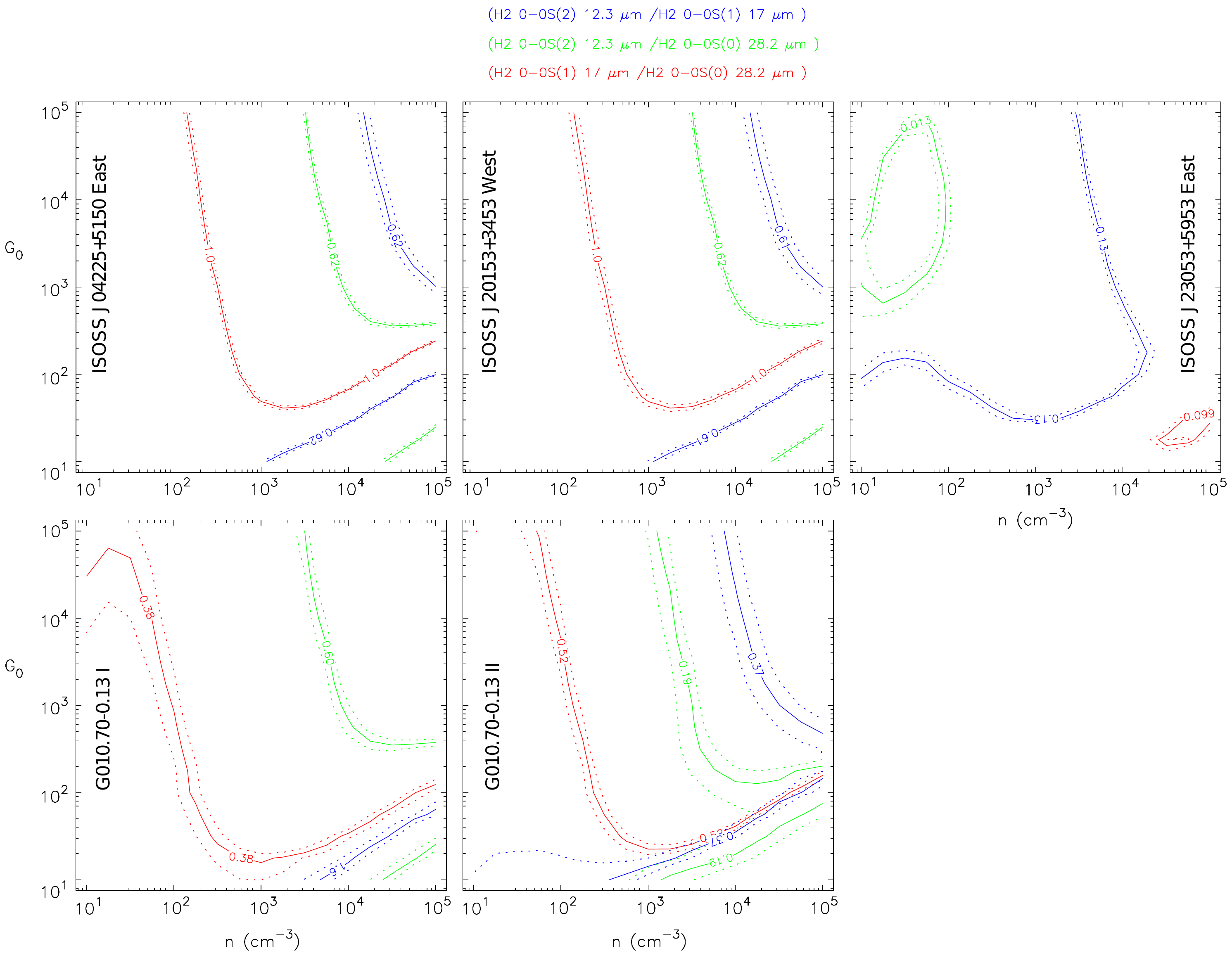}
	\caption{Best fitting PDR models for certain H$_2$ line ratios from \citet{Kaufman06} are shown. The line fluxes are obtained from the high resolution orders. For visualization we used the PDR toolbox  (\texttt{http://dustem.astro.umd.edu/pdrt/)}. To show the contrast, ISOSS~J23063$+$5963 is plotted as a region which does not host a PDR.} 
	\label{fig:PDRmodel}
\end{figure*}

ISOSS J23053$+$5953 appears as a deeply embedded source. The extinction, calculated from the 9.7~$\mu$m silicate feature, is highest in our sample (see Table \ref{tab:tau}).
The bright 11.2~$\mu$m PAH feature is spatially extended over more than 1' with almost constant line strength, as shown in the spatial line fits (Figure \ref{fig:spatial_line_23}). The spectrum of ISOSS J23053$+$5953~East is dominated by neutral PAHs. 
No spectral indications for a PDR such as, e.g., strong cationic PAH bands were found in this region.  
Furthermore, by fitting the H$_2$ line ratios  to the PDR models by \citet{Kaufman06} no correlation between density $n$ and FUV intensity $G_0$ modeled for different line ratios was found (see Figure \ref{fig:PDRmodel}).

Therefore the observed forbidden lines of \FeII, \SiII and \SI do not originate from the hot layers of a PDR. These lines can be formed in the dense hot post-shocked gas of J-shocks \citep{Hollenbach89}. Furthermore, the observed \SI line is never strong in PDRs \citep{Kaufman06}, but can be observed as a velocity independent emission line in J-shocks \citep{Hollenbach89}.
The presence of higher H$_2$ rotation lines is typical for the post-shock gas phase \citep{Lahuis06}. Therefore, a possible interpretation of the spatial distribution of the $S(5)$ H$_2$ line is that it represents the distribution of the cooling shocked gas, (the $S(5)$ line was only detected sidewards of the MIPS 24~$\mu$m peak position). However, fluorescence pumping cannot be completely ruled out as an excitation mechanism.
 The observed ``green and fuzzy'' features in the IRAC bands and narrow band imaging ($\lambda=2.122$~$\mu$m, H$_2$, $S(1)$, 1-0) reveal the presence of excited H$_2$. This supports the hypothesis of molecular hydrogen emissions from a post-shock relaxation region. The 6.22~$\mu$m PAH band is detected towards most of the positions where the $S(5)$ H$_2$ line was observed.  Therefore, the excitation mechanism could be the same for this particular PAH as for the $S(5)$ H$_2$ lines.  
No molecular emission from CO, H$_2$O, or OH, indicators for C-shocks, was found. Also the line H$_2$ ratios disagree with the results of \citet{Kaufman96}.
The symmetric decrease of the $S(3)$ line fluxes around the
source position indicates that a predominant fraction of the gas is thermally excited by the central source. The cooler gas traced by the $S(1)$ line is extended over more than 2.5$'$. Absorption bands of CO$_2$ ice and amorphous silicates indicate the presence of an outer ridge of cold material facing the observer.


\subsection{ISOSS J18364$-$0221 West - An isolated, icy source} \label{sec:18W}
\begin{figure}[h]
 	\centering
	\plotone{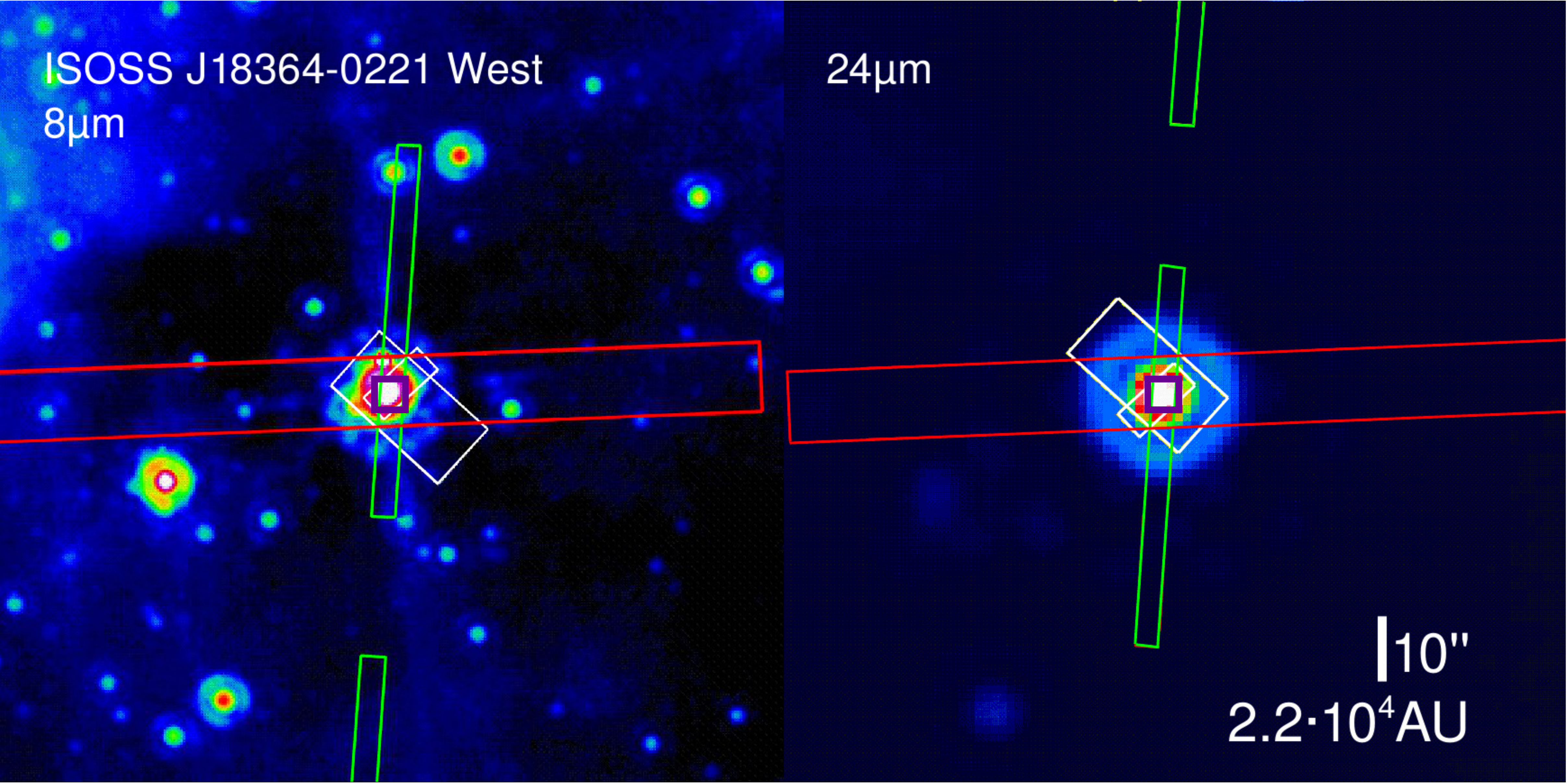}
	\caption{Overlay of the IRS slits for ISOSS J18364$-$0221 West. Notations are the same as those for Figure~\ref{fig:slit_20EW}.}
	\label{fig:slit_18W}
\end{figure}

\placefigure{fig:slit_18W}

\begin{figure}[h]
	\centering
	\plotone{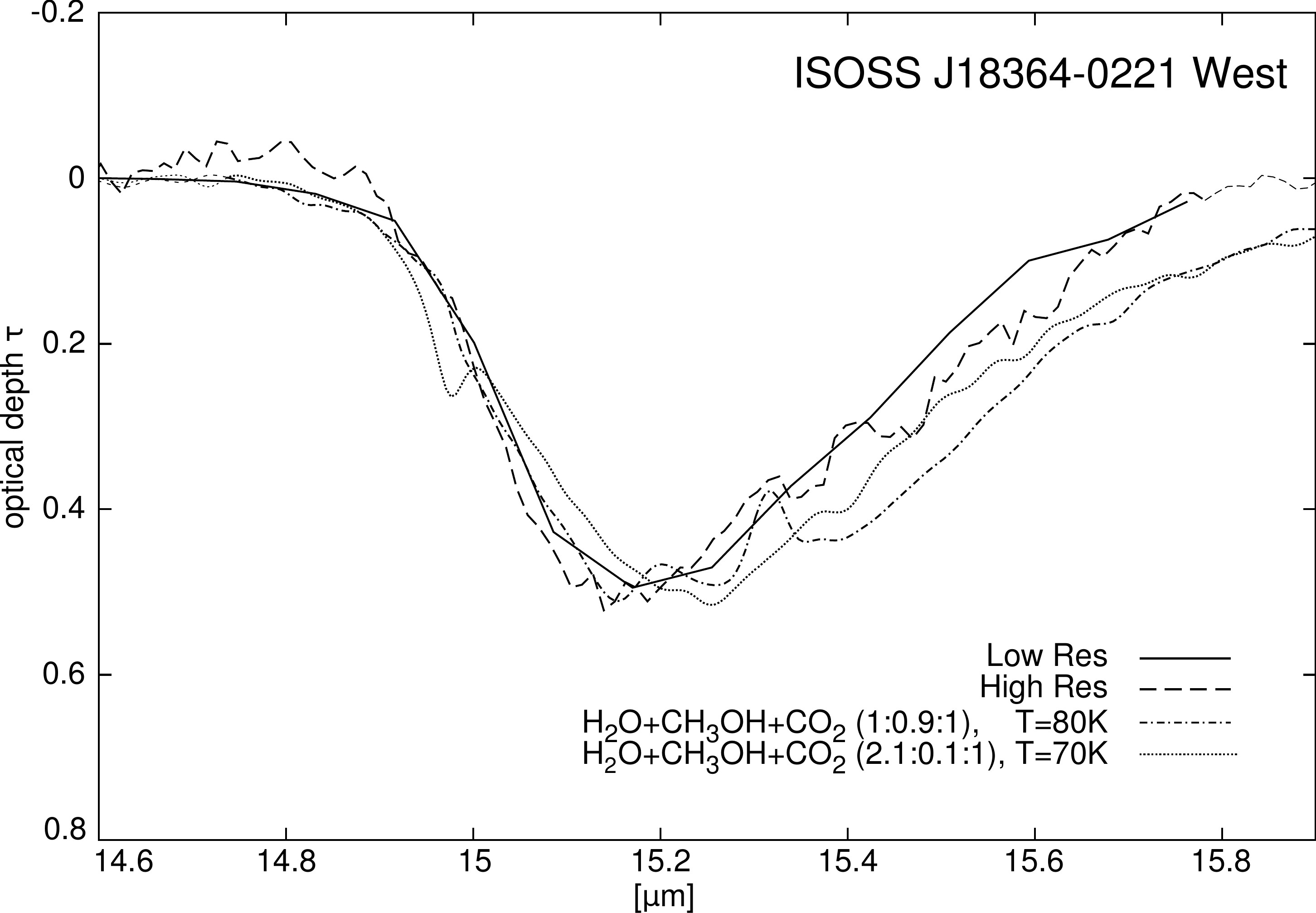}
	\caption{Optical depths of the 15.2~$\mu$m ice feature for ISOSS J18364-0221 West. The solid line represents the low-resolution spectrum. The long dashed line represents the high-resolution spectrum. The two other lines represent laboratory absorbance data for an ice mixture of methanol, water, and carbon dioxide for temperatures of 70 and 80~K \citep{White09}.}
	\label{fig:CO2_18W}
\end{figure}

The ISOSS J18364$-$0221 region 2.2~kpc away has been studied by \citet{Birkmann06}.
These observations show a large scale submillimeter clump with two separate cores in the SCUBA bands. For the western component a mass of $M=280^{+75}_{-60}\;M_{\sun}$ and a temperature $T_d=12.8$~K were estimated.

The IRAC observations show a bright compact object, some weak extended emission features and a faint point source eastward on the first airy ring. The bright point source on the northern edge of the airy ring of the  central source must be treated as an artifact since it does not show the characteristic IRAC PSF. In the MIPS bands a single point source appears centered on the central bright IRAC source. The cross dispersion profile appears to be oblate and broadened, hence the target is consistent with the IRAC observations of a compact, but extended, source.
The spectroscopy shows several ice absorption features and a broad silicate absorption at 9.7~$\mu$m. 

We used the IRAC and MIPS luminosities and the silicate extinction (Table \ref{tab:tau}) with the SED fitting tool from \citet{Robitaille07} to obtain an estimate for the source parameters. For ISOSS J18364$-$0221 West, the modeling favors masses below 2~$M_\odot$ and accretion rates between $10^{-5}\;-\; 3\times 10^{-4}\;M_\odot\textrm{yr}^{-1}$. 
The modeled envelope and ambient masses are proportional to the accretion rates and range from $0.1-10 M_\odot$. 
This is consistent with the scenario of a deeply embedded young low-mass protostar which could presumably evolve into a low-to-intermediate mass star. \citet{Birkmann06} had already classified this object as a low mass class I object based on the NIR colors.

The absorption profile for the 15.2~$\mu$m ice feature cannot be explained by pure CO$_2$. We compared the profile of ISOSS J18364$-$0221 West with different absorption profiles for H$_2$0+CH$_3$OH+CO$_2$ mixtures from \citet{White09}. The best fit was found for ice mixtures heated to 70-80~K (Figure \ref{fig:CO2_18W}). Therefore, the protostar is in fact the heating source of the ice, and not a background object.


\subsection{Young embedded IRDC sources} \label{sec:Ygalactic}
\begin{figure}
 	\centering
 	\plotone{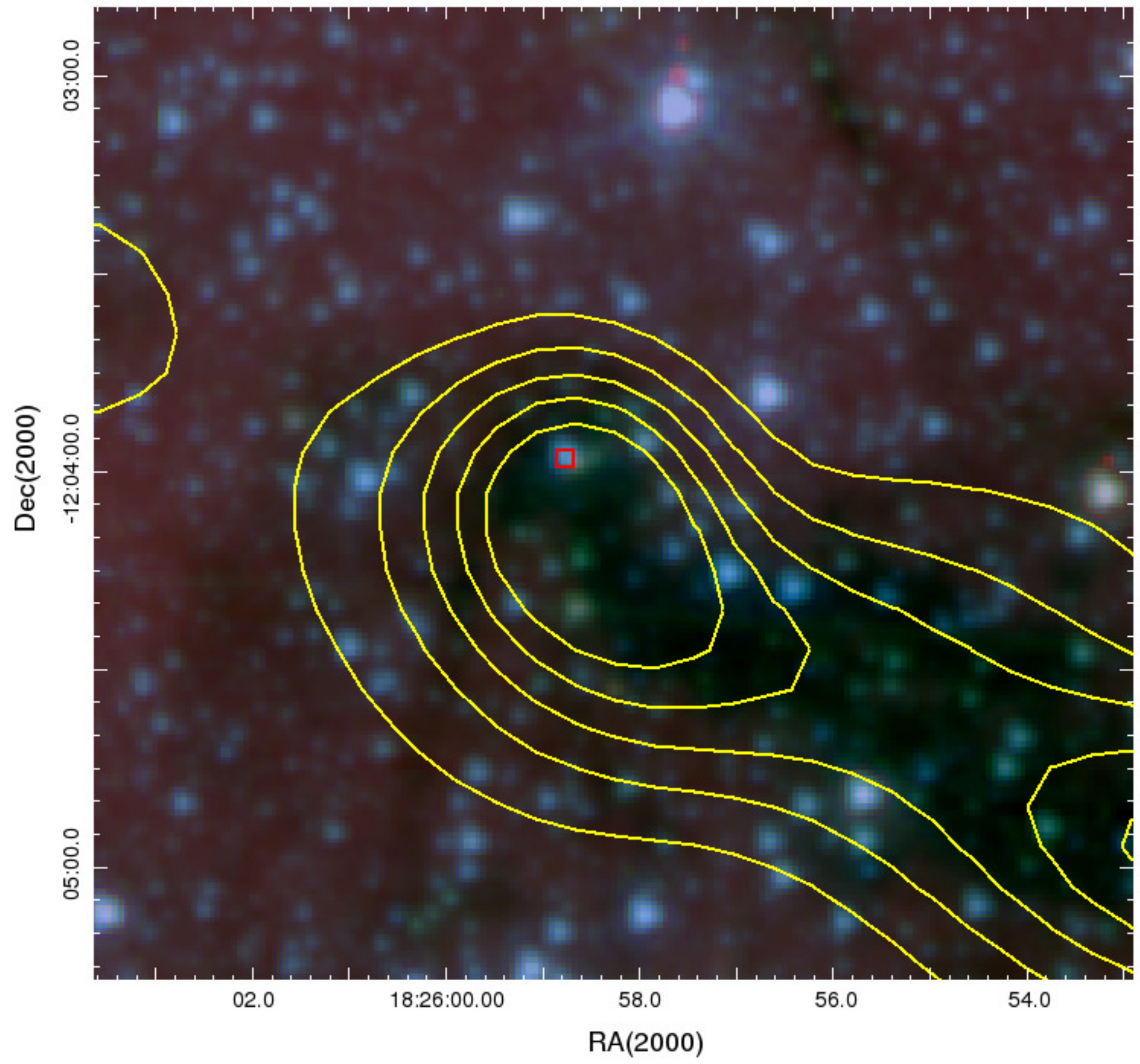}
	\caption{IRAC composite image for G019.27+0.07. The 3.6~$\mu$m channel is colorized in blue, 4.5~$\mu$m in green and 8.0~$\mu$m in red.  The peak position of the MIPS points source at 24~$\mu$m is indicated by the red box points. The IRS slit is centered on the position. The SCUBA observation at 850~$\mu$m is shown as yellow contours.}
	
 \label{fig:map_G19}
\end{figure}

G019.27+0.07 appears in the IRAC bands as an isolated point source at the northeast extension of an IRDCfilament. The background outside the filament and the shape of the filament change depending on the wavelength. This makes the spectral extraction for this source quite difficult. Therefore the source spectrum with a polynomial background fit and the nodded background spectrum is shown in Figure \ref{fig:spec_G19}. The 5.8$\mu$m and 8$\mu$m bands show the extinction of the IRDC against the galactic PAH background. 
The IRAC source is also detected as a point source in MIPS24/70. An extended submillimeter clump is observed at the position of the IRAC/MIPS source as denoted by the contours in Figure \ref{fig:map_G19}. The clump can also be seen at 1.2mm, and the core mass was calculated as $113M$\sun  ~\citep{Rathborne06}. H$_2$O and thermal CH$_3$OH maser emissions were observed with the Green-Bank-Telescope by \citet{Chambers09}. The later maser emission indicates a more evolved evolutionary stage.
The background spectrum of G019.27+0.07 reveals a $\mathcal{A}$ spectrum. These PAH features are not detected toward the source spectrum, hence the UV excitation source can be external. Molecular hydrogen emissions are found in the source and background spectra. This is consistent with a slightly extended blob of H$_2$ observed in the IRAC band at 4.5~$\mu$m.

G034.43+0.24 is an isolated source found in MIPS and IRAC images. The peak of the submillimeter clump at 850~$\mu$m is centered on the 24~$\mu$m source position. The extended lobe in the 3.6~$\mu$m map of molecular hydrogen lines could be excited by an outflow. The same absorption feature as that in ISOSS J18364$-$0221 West is observed. This supports the hypothesis of a young, deeply-embedded object with outflows. 
For the remaining targets embedded in IRDCs, no extended hydrogen features were resolved on the IRAC maps. However, molecular hydrogen emission was detected in G028.04-0.46 and G053.25+0.04~I+II.

\subsection{Comparison with Orion IRc2}
Many of the general features observed in the \textit{ISO} spectra of our sample are also observed in the spectrum of Orion IRc2 (see Table \ref{tab:Lbol}).

The 9.7~$\mu$m silicate absorption feature can be found in almost every source. As in IRc2, we find ice features of water, methane and CO$_2$. NH$^+_4$ in the gas phase appears in absorption. Several molecular features from IRc2, e.g., SO$_2$, C$_2$H$_2$, and CH$_4$ (in the gas phase), are not present in our sample at the IRS resolution and sensitivity.
Most sources show a wide variety of PAH bands. Since the times of the \textit{ISO} observations of IRc2, there has been remarkable progress in the understanding of PAH composition \citep{Tielens08}. We constrained the physical properties of the far-UV for several sources from the PAH bands at 6.8, 7.7, and 8.6~$\mu$m. 
Due to the lower spectral resolution of \textit{Spitzer} IRS compared to \textit{ISO}/SWS, features like H$_2$O in the gas phase can not be observed in our sample.  
Since the spectra of IRS are taken longward of 5.5~$\mu$m, only the 0-0 $S(0)$-$S(7)$ lines can be observed. The $Q$ and $O$ modes and the higher orders for $S$ remain inaccessible. It is difficult to distinguish between the different excitation mechanisms for molecular hydrogen, as was done for the Orion peaks \citep{Rosenthal00}, because we are missing these orders at shorter wavelengths. However, the spatial distributions of the $S(5)$ and $S(3)$ lines enable us to discriminate between thermal excitation and shock excitation. For some of the sources in our sample the \NeII line is observed in our sample as well, and can be correlated to a PDR and/or J-shock excitation. In contrast to the Orion starforming region we did not find any indications for C-shocks. 
All spectra in our sample show only a fraction of Orions IRc2 characteristics. Either they show deeply embedded young sources with strong silicate and ice absorption bands, or more evolved sources with indications of PDRs and strong PAH bands below 9~$\mu$m.

\section{Summary} \label{sec:Summary}
We used the IRS on board of the \textit{Spitzer Space Telescope} to observe a sample of 14 intermediate-mass YSOs in different evolutionary stages. We present low- and-high resolution mid-infrared spectra of these sources.
The IRS spectra demonstrated how difficult it is to disentangle the complexity of such starforming regions, because the IRS beam covers multiple physical components. However, with the help of previous multi-wavelength studies we could shed light on several physical components and properties. In contrast to previous spectroscopic \textit{ISO} observations, we derived information about the spatial extent of lines using the IRS slit spectroscopy.
The spatial distribution of selected PAH and H$_2$ lines in comparison with previous imaging observation can shed light not only on the YSOs but also on their vicinity.
The infrared spectroscopy shows two different kinds of sources:
\begin{itemize}
 \item Young isolated sources, which are still embedded in an envelope of cold dust and ice. These sources tend to still have low to intermediate masses with high accretion rates. 
Only neutral background PAHs are found for these sources. 
 Several different absorption features, such as NH$^+_4$ (6.75~$\mu$m), water, methane, and CO$_2$ ice, can be found for these sources.
CO$_2$ ice can also be found for the more evolved sources too. For ISOSS~J23053+5953~East the same indications for J-shocks are found as for the more evolved source ISOSS~J20153$+$3453~West (see below).

 \item The second group of sources is more evolved. We found H$_2$ pure rotational lines, atomic fine structure lines such as \FeII, \SiII and several PAH bands. The presence of \SiII, \FeII and H$_2$ lines can be explained by J-shocks. For some sources \NeII indicates high-velocity shocks. The line ratios for \NeII and H$_2$, \SiII, \FeII agree with the predictions for J-shocks. Shock excitation might also explain the spatial distribution of certain H$_2$ lines for ISOSS~J20153$+$3453~West. There are no indications for C-shocks in our sample. For some sources a class $\mathcal{A}$ PAH spectrum was detected with strong PAH bands below 9~$\mu$m. All these sources have indications for a PDR. Although the contribution of shock excitations to the H$_2$ line emission is not known, PDR models \citep{Kaufman06} favor higher gas densities ($\geq10^5\;\textrm{cm}^{-3}$) and UV field strengths between $10^2$ and $10^4\;\textrm{G}_0$.
\end{itemize}

\acknowledgements
We thank Zoltan Balog for his help with the reduction of IRAC/GLIMPSE data for the IRDC sources. We appreciate the help of Bernhard Sturm with the polynomial background fits for some of the low-resolution spectra. This work is based in part on observations made with the \textit{Spitzer Space Telescope}, which is operated by the Jet Propulsion Laboratory, California Institute of Technology, under a contract with NASA.

\end{document}